%% file: main.tex
\documentclass[journal=jacsat,manuscript=article]{achemso}
\usepackage{xr-hyper}
\usepackage{siunitx}
\usepackage{graphicx} 
\usepackage[utf8]{inputenc}
\usepackage[T1]{fontenc}
\usepackage{amsmath}
\usepackage{amsfonts}
\usepackage{amssymb}
\usepackage{array}
\usepackage[version=3]{mhchem} % Formula subscripts using \ce{}
\usepackage{braket}
\usepackage{bm}
\usepackage{xspace}
\usepackage{multirow}

\usepackage{cleveref}
\usepackage{titlesec}

\usepackage{xspace} 
\usepackage{caption}
\usepackage{subcaption}
\usepackage{tikz}
\usetikzlibrary{arrows, arrows.meta}
\usepackage{comment}
\usepackage{newtxtext}
\usepackage{mathtools}
\setkeys{acs}{articletitle = true}

\DeclareUnicodeCharacter{0301}{\'{e}}

\makeatletter
\newcommand*{\addFileDependency}[1]{
 \typeout{(#1)}
 \@addtofilelist{#1}
 \IfFileExists{#1}{}{\typeout{No file #1.}}
}
\makeatother

\newcommand*{\myexternaldocument}[1]{
 \externaldocument{#1}
 \addFileDependency{#1.tex}
 \addFileDependency{#1.aux}
}
\myexternaldocument{SI}

\author{Chiara Sepali}
\affiliation{Scuola Normale Superiore, Piazza dei Cavalieri 7, I-56126 Pisa, Italy}
\alsoaffiliation{C.S. and P.D.C.  contributed equally to this work}
\author{Pietro Dal Cin}
\affiliation{Scuola Normale Superiore, Piazza dei Cavalieri 7, I-56126 Pisa, Italy}
\alsoaffiliation{C.S. and P.D.C.  contributed equally to this work}
\author{Alicia M. Kirk}
\affiliation{Scuola Normale Superiore, Piazza dei Cavalieri 7, I-56126 Pisa, Italy}
\author{Sonia Coriani}
\affiliation{DTU Chemistry, Technical University of Denmark, Kemitorvet Bldg. 207, 2800 Kongens Lyngby, Denmark}
\author{Chiara Cappelli}
\affiliation{Scuola Normale Superiore, Piazza dei Cavalieri 7, I-56126 Pisa, Italy}
\email{chiara.cappelli@sns.it}

\title{Multireference Polarizable Embedding for Core-Level Spectroscopy in Aqueous Solution: XPS and XAS of Glycine}

\abbreviations{IR,NMR,UV}
\begin{document}

\begin{abstract}

Core-level spectra in solution are governed by the interplay of core-hole relaxation, electron correlation, specific solute--solvent interactions, and environmental polarization. We report the first application of RASSCF--RASPT2 combined with the fluctuating-charges polarizable embedding to X-ray photoelectron 
(XPS) and X-ray absorption (XAS) spectroscopy in aqueous solution. The protocol couples a multireference treatment of core-ionized and core-excited states with molecular-dynamics sampling, and is assessed for the carbon and nitrogen K-edges of glycine in the gas phase and in water. The calculated spectra are compared with experiment and with fixed-charges electrostatic embedding. Gas-phase ionization and absorption features are accurately reproduced. In solution, configurational averaging captures inhomogeneous broadening, while XPS peak positions, relative intensities, and widths depend strongly on the solvent parametrization. The best-performing polarizable model improves the spectral profile, although aqueous ionization energies remain overestimated. For XAS, 
the carbon K-edge is reproduced close to the experimental energy scale, whereas the nitrogen K-edge line shape is recovered after rigid alignment. These results support multireference polarizable embedding as a promising framework for core-level spectroscopy in solution.

\end{abstract}

%%%%%%%%%%%%%%%%%%%%%%%%%%%%%%%%%%%%%%%%%%%%%%%%%%%%%%%%%%%%%%%%%%%%%
%% Start the main part of the manuscript here.
%%%%%%%%%%%%%%%%%%%%%%%%%%%%%%%%%%%%%%%%%%%%%%%%%%%%%%%%%%%%%%%%%%%%%

\newpage

\section*{Introduction}

Core-level spectroscopies provide element- and site-selective probes of molecular electronic structure and have become central tools for investigating chemical bonding, charge redistribution, and ultrafast molecular dynamics.
\cite{Stohr1992NEXAFS,norman2018simulating,kraus2018ultrafast,Bhattacherjee2018}
At the same time, advances in bright synchrotron sources and X-ray free-electron lasers have extended the scope of these techniques to femtosecond and even sub-femtosecond time scales, enabling direct observation of electronic and nuclear dynamics through X-ray absorption spectroscopy (XAS), X-ray photoelectron spectroscopy (XPS), ultrafast scattering, and nonlinear X-ray spectroscopies.
\cite{kraus2018ultrafast,young2018roadmap,foglia2018first,mcneil2010x,
goulielmakis2010real,attar2017femtosecond,bressler2009femtosecond,
wolf2017probing,van2009structural}
The interpretation of XPS and XAS signals based only on experimental findings is challenging. Therefore, they increasingly rely on accurate computational protocols capable of assigning experimental features and connecting them to the local electronic structure of the probed atomic centers. 
\cite{coriani2012coupled,coriani2012asymmetric,fransson2013carbon,
wenzel2014calculating,wenzel2015analysis,brabec2012communication,
sen2013study,peng2015energy,coriani2015comm,
vidal2019fcCVSEOM,vidal2020equation,
park2022mrtddft,norman2018simulating,bokarev2020theoretical,montorsi2022soft,alias2026benchmarking}

Despite the localized character of core orbitals, the associated spectroscopic response is not local from a computational perspective. The creation of a core hole induces substantial orbital relaxation and redistribution of the valence density, while core-excited states may involve diffuse, strongly mixed, or multiply excited configurations.~\cite{norman2018simulating}  In condensed phases, these effects are further coupled to specific intermolecular interactions and to the response of the surrounding environment. Hydrogen bonding, structural fluctuations, and differential solvent stabilization of the initial and final states can produce sizable energy shifts, redistribute spectral intensity, and generate pronounced inhomogeneous broadening. \cite{warshel1976theoretical,warshel1972calculation,tomasi2005quantum,
giovannini2020molecular,smith2017soft,weinhardt2013rixs,
lange2013electronic}
XAS is particularly sensitive to local hydration because core excitations often involve diffuse and geometry-dependent unoccupied orbitals.
\cite{smith2017soft,wernet2004structure}
Such effects have been documented for amino acids, peptides, and proteins, including pH-dependent nitrogen K-edge signatures of aqueous amino acids and alkylamines.
\cite{messer2005local,schwartz2010investigation,messer2005ph,
ekimova2018soft}
Predictive simulations of solution-phase core spectra therefore require a balanced treatment of both the molecular electronic structure and solvent effects.

Several families of electronic-structure methods have been developed for the calculation of core-ionization and core-excitation energies.
\cite{norman2018simulating,bokarev2020theoretical,leetmaa2010theoretical}
Density-functional theory (DFT)  and time-dependent (TD) DFT remain attractive for large systems because of their favorable 
balance between accuracy and computational cost. Specialized approaches, including transition-potential, maximum-overlap, and \textit{hole--hole} Tamm--Dancoff formulations, extend their applicability to high-energy states and states with non-negligible multiple-excitation character, although important limitations remain.
\cite{ehlert2018efficient,attar2017femtosecond,yu2020ab,
bokarev2020theoretical,park2022mrtddft}
Core excitations can also be selectively targeted within TDDFT through the core--valence separation (CVS) approximation, which restricts the excitation manifold to configurations involving at least one core orbital.
\cite{cederbaum1980cvs,stener2003cvs,fransson2021xaboom,coriani2015comm}

Wave-function-based approaches, such as algebraic-diagrammatic construction methods, linear-response coupled cluster, and equation-of-motion coupled cluster, 
especially when coupled to the CVS approximation, generally provide a more systematic treatment of core-hole relaxation and dynamic electron correlation. \cite{schirmer1982beyond,coriani2015comm,vidal2019fcCVSEOM,vidal2020equation,
fransson2021xaboom,barth1985theoretical,trofimov2000core,
peng2015energy}
Their predominantly single-reference character may nevertheless become problematic when the target states possess substantial double- or higher-excitation character, strong configurational mixing, or near-degeneracies between valence and core-excited configurations. \cite{yu2020ab,
vidal2020xps,tsuru2022,chang2025pyrazine}
%vidal2020equation,

Multiconfigurational approaches provide a natural framework for these situations. Restricted active space self-consistent field theory (RASSCF)\cite{olsen1988determinant,malmqvist1990restricted} and its second-order perturbative extension, RASPT2,
\cite{malmqvist2008restricted} allow core orbitals, valence orbitals, and selected virtual orbitals to be treated within a common correlated wave function. These methods have been successfully applied to K- and L-edge spectra of small organic molecules,\cite{schnorr2019tracing,nenov2019x} molecular ions,\cite{lindblad2020x,couto2020carbon} and transition-metal complexes. \cite{pinjari2014restricted,pinjari2016cost,preusse2016towards,
guo2018fingerprinting}
The incorporation of CVS-based core-targeting strategies into the RASSCF/RASPT2 framework enables the selective treatment of valence, single- and multiple-core excitations, mixed core--valence states, and core-ionized manifolds. \cite{delcey2019efficient}
This development has facilitated simulations of time-resolved XAS signals \cite{hua2019transient,northey2020ultrafast,segatta2020exploring} and provided access to the multiple-core states required for nonlinear X-ray spectroscopies. \cite{nenov2019x,gu2021manipulating} Recent benchmarks have further assessed the performance of RASPT2 and other multireference approaches for K-edge absorption energies across broad molecular test sets.\cite{alias2026benchmarking} Nevertheless, most multireference applications have so far focused on isolated molecules, whereas their integration with realistic descriptions of condensed-phase environments remains comparatively underdeveloped.

Environmental effects can be incorporated through focused multiscale models, in which the molecular subsystem of interest (the solute) is treated at a high level of quantum-mechanical theory while the surrounding environment is represented using a computationally less demanding description. \cite{warshel1972calculation,warshel1976theoretical,senn2009qm, lin2007qm,giovannini2020molecular,gomez2022multiple, bondanza2020polarizable} Among these approaches, quantum mechanics/molecular mechanics (QM/MM) methods describe the solvent explicitly through an atomistic force field and can therefore capture local and directional interactions such as hydrogen bonding. \cite{warshel1972calculation,warshel1976theoretical,senn2009qm, lin2007qm,morzan2018spectroscopy,giovannini2020molecular,
bondanza2020polarizable,orozco2000theoretical}
In conventional electrostatic embedding, the environment is represented by fixed atom-centered charges whose electrostatic potential enters the quantum-mechanical Hamiltonian.\cite{senn2009qm,lin2007qm} Such models allow the solvent field to polarize the quantum-mechanical subsystem but neglect the reciprocal response of the environment to changes in the solute electron density. This approximation may be particularly restrictive for core spectroscopy because the neutral, core-ionized, and core-excited states can generate markedly different electrostatic potentials and may therefore polarize the solvent to different extents.

Polarizable embedding models restore this mutual response by allowing selected environmental degrees of freedom to adapt self-consistently to the quantum-mechanical electron density. \cite{senn2009qm,lin2007qm} Different formulations have been proposed, based on induced dipoles, distributed polarizabilities, multipoles, or fluctuating charges (FQ). \cite{loco2016qm,thole1981molecular,curutchet2009electronic,olsen2011molecular,steindal2011excitation,giovannini2019polarizable,boulanger2012solvent,rick1994dynamical,thompson1995excited,dziedzic2016tinktep,cappelli2016integrated,giovannini2020molecular,gomez2022multiple}

The FQ model has recently been combined with multiconfigurational self-consistent-field wave functions, resulting in a fully variational MCSCF/FQ polarizable embedding framework applicable to both ground and  electronically excited states. \cite{sepali2024fully} Its potential for core-level spectroscopy, however, has not yet been systematically investigated. More generally, applications combining multireference wave functions, atomistic polarizable embedding, and extensive configurational sampling remain scarce for XPS and XAS in solution. \cite{larsson2024simulating,ross2018comprehensive}

In this work, we present the first application of the RASSCF–RASPT2/FQ implementation of the MCSCF/FQ framework to the simulation of both XPS and XAS spectra in aqueous solution. Glycine is selected as a stringent benchmark. Its small size permits a detailed assessment of active-space and correlation effects, whereas the transition from an ensemble of neutral gas-phase conformers to a strongly hydrated zwitterionic structure in water provides a demanding test of the environmental description. In particular, the charged ammonium and carboxylate groups generate a structured and anisotropic hydration environment, making the resulting spectra sensitive to local hydrogen bonding, solvent fluctuations, and mutual polarization. Carbon and nitrogen K-edge XPS and XAS spectra 
in solution are obtained by averaging over configurations extracted from classical MD simulations. The polarizable RASSCF--RASPT2/FQ results are compared with gas phase spectra, a non-polarizable fixed-charge embedding and with available experimental data. Different active spaces and two FQ parametrizations are examined to determine how the electronic-structure treatment and the environmental response affect absolute energies, peak separations, relative intensities, and inhomogeneous broadening. The role of explicit core--valence correlation in the perturbative treatment is also assessed. Globally, the results of the present study establish the capabilities and current limitations of multireference polarizable embedding for core-level spectroscopy and provide a foundation for its application to more complex solvated molecular systems.

\section{Theory}
This section provides an overview of the MCSCF/FQ framework, following Ref.~\citenum{sepali2024fully}. The computational protocols used to compute XAS and XPS spectra in the gas phase and in aqueous solution are then described, employing RASSCF–RASPT2 for the gas-phase calculations and RASSCF–RASPT2/FQ in solution. 

\subsection{The MCSCF/FQ approach}\label{sec:coupling}

The FQ force field \cite{rick1994dynamical,rick1995fluctuating,cappelli2016integrated} assigns a variable charge $q$ to each atom of the classical, atomistic region (here, the aqueous solvent). These charges are not fixed but respond self-consistently to the electrostatic potential generated by the surrounding environment. Charge redistribution follows the electronegativity equalization principle (EEP) \cite{sanderson1951interpretation}, according to which the instantaneous atomic electronegativities $\chi_{i\alpha}$ equilibrate at convergence. An equivalent convenient formulation expresses the fluctuating charges as the constrained minimum of the functional: \cite{cappelli2016integrated,giovannini2020molecular}
\begin{equation}\label{eq:energy_fq}
E_{\text{FQ}} = 
\sum_{i \alpha} q_{i \alpha} \chi_{i \alpha}
+ \frac{1}{2} 
\sum_{i \alpha} \sum_{j \beta} 
q_{i \alpha} T_{i \alpha, j \beta}^{qq} q_{j \beta}
+ \sum_{\alpha} 
\lambda_\alpha 
\left( \sum_i q_{i \alpha} - Q_{\alpha} \right)
\end{equation}
where indices $i,j$ label atoms and $\alpha,\beta$ label molecules.  
A set of Lagrange multipliers $\lambda_\alpha$ enforces conservation of each molecule’s total charge $Q_\alpha$, thereby preventing unphysical intermolecular charge transfer \cite{chen2009dissociation}.  
The kernel $T_{i\alpha,j\beta}^{qq}$ describes charge–charge interactions and is modeled using the Ohno ansatz \cite{ohno1964some} and, to prevent the polarization catastrophe \cite{thole1981molecular}, the diagonal elements are linked to the atomic chemical hardness $\eta_{i\alpha}$.
In practice, the FQ model is defined by two atomic parameters: the electronegativity $\chi_{i\alpha}$ and the chemical hardness $\eta_{i\alpha}$.

FQ can be coupled to a QM description of a portion of the total system (the solute in the present case) within a  QM/MM-like formalism. The total energy of the MCSCF/FQ system reads: \cite{sepali2024fully}

\begin{equation}\label{eq:total-energy}
E_{\mathrm{tot}} = E_{\mathrm{MCSCF}} + E_{\mathrm{FQ}}
+ E_{\mathrm{MCSCF/FQ}}^{\mathrm{int}}
\end{equation}
where $E_{\mathrm{MCSCF}}$ is the multiconfigurational electronic energy for the isolated solute,\cite{roos1980complete} and $E_\textrm{FQ}$ the energy of the classical FQ portion.\cite{sepali2024fully} 

The coupling term is given by: \cite{cappelli2016integrated,sepali2024fully}
\begin{equation}\label{eq:casscf-fq}
E^\mathrm{int}_{\text{MCSCF/FQ}} = \sum_{i\alpha} q_{i\alpha} \, V^{i\alpha}(\mathbf{D})
\end{equation}
where $V^{i\alpha}(\mathbf{D})$ is the electrostatic potential generated by the QM density ($\mathbf D$) at the position of atom $i$ in molecule $\alpha$. 
In the MCSCF/FQ scheme, the effect of the polarizable environment enters the electronic structure calculation through a one-electron term added to the Hamiltonian:\cite{sepali2024fully}
\begin{equation}\label{eq:ham_eff}
\hat{H} = \sum_{pq} \left[ h_{pq} + \sum_{i\alpha} q_{i\alpha} V^{i\alpha}_{pq} \right] \hat{E}_{pq} +  \frac{1}{2} \sum_{pqrs} g_{pqrs} \left( \hat{E}_{pq} \hat{E}_{rs} - \delta_{rq}\hat{E}_{ps} \right)
\end{equation}
with elements $V^{i\alpha}_{pq}$ given by:
\begin{equation}
    V^{i\alpha}_{pq} =  -\Braket{\phi_p | \frac{1}{\left|\boldsymbol{r}_{i\alpha} - \boldsymbol{r}\right|} |\phi_q}~.
\end{equation}
The total energy functional is minimized with respect to both MCSCF parameters and FQ charges until full self-consistency is reached \cite{sepali2024fully}. Minimization with respect to the FQ degrees of freedom yields the following linear system:
\begin{equation}\label{eq:lin_sys}
\left(
\begin{array}{cc} 
\mathbf{T}^{\textit{qq}} & \mathbf{1}_{\lambda} \\ 
\mathbf{1}^{\dagger}_{\lambda} & \mathbf{0}
\end{array}
\right)
\begin{pmatrix}
\mathbf{q}(\mathbf{D}) \\ \bm{\lambda}
\end{pmatrix}
=
\begin{pmatrix}
-\bm{\chi} \\ \mathbf{Q}_{\alpha}
\end{pmatrix}
+
\begin{pmatrix}
-\mathbf{V}(\mathbf{D}) \\ \mathbf{0}
\end{pmatrix}
\end{equation}
explicitly showing that the QM density polarizes the FQ region.  
The resulting FQ charges, in turn, modify the MCSCF Hamiltonian through \cref{eq:ham_eff}, thereby ensuring fully mutual, self-consistent solute–solvent polarization. The formulation applies equally to state specific (SS) and state average (SA) multiconfigurational schemes.  

Among the MCSCF methods, CASSCF \cite{roos1980complete,roos1980complete1} performs a full-CI treatment within the chosen active space, whereas RASSCF \cite{malmqvist1990restricted} partitions the active orbitals into three subspaces—RAS1, RAS2, and RAS3—allowing controlled restrictions on the number of holes and particles. RAS1 contains up to a given maximum number of holes; in RAS2 all possible distributions of electrons are allowed;
and RAS3 contains up to a given maximum number of electrons. This flexibility is especially valuable when treating core orbitals or large active spaces.
Dynamic correlation can be included via second-order perturbation theory, leading to CASPT2 \cite{andersson1990second,andersson1992second} or RASPT2 \cite{malmqvist2008restricted,sauri2011multiconfigurational}.  
In the present embedding framework, the perturbative step is carried out using the converged FQ charges from the MCSCF calculation, without further polarization of the environment.

Several multistate PT2 formalisms are available.  
MS-CASPT2(RASPT2) accounts for interactions among quasi-degenerate states through a multistate effective Hamiltonian \cite{finley1998multi}.  
XMS-CASPT2(RASPT2) enforces invariance with respect to rotations within the reference space, improving robustness in the presence of strong state mixing \cite{granovsky2011extended,shiozaki2011communication}.  
RMS-CASPT2(RASPT2) further stabilizes the description of state crossings by introducing a rotated reference space \cite{battaglia2021role}.  
All these formalisms can be used consistently within the MCSCF/FQ embedding framework.

\section{Protocol for gas-phase and condensed phase XPS and XAS}\label{sec:xps_xas}

\begin{figure}[b!]
    \centering
    \includegraphics[width=0.8\linewidth]{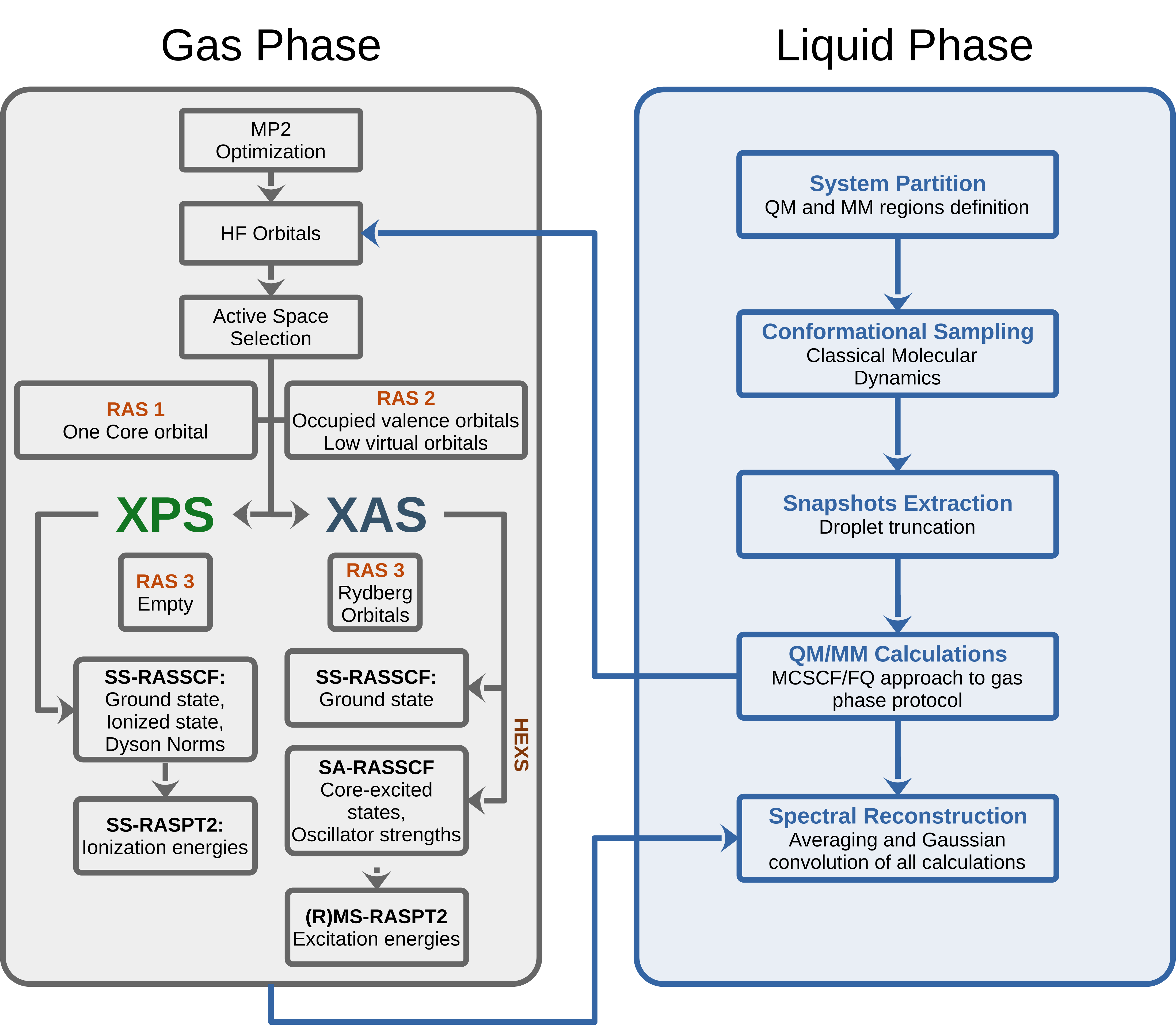}
    \caption{Workflow of gas-phase and solution phase protocols for the computation of XPS and XAS}
    \label{fig:protocol}
\end{figure}

The RASSCF-RASPT2 method is employed to compute gas-phase XPS and XAS, explicitly targeting core-ionized and core-excited states, in line with previous applications to isolated molecules. \cite{montorsi2022soft}
The gas-phase protocol (see Fig.\ref{fig:protocol}) begins with geometry optimization and the computation of Hartree-Fock (HF) orbitals, which are then used to define a chemically meaningful active space.

For XPS, the targeted core orbital is assigned to RAS1 and kept frozen, while the valence orbitals together with a few low-lying virtuals are included in RAS2; RAS3 remains empty. This setup is generally robust and ensures stable convergence of the RASSCF optimization.
For XAS, the same core orbital is placed in RAS1, valence occupied orbitals fill RAS2, and RAS3 includes the valence virtual orbital as well as a compact set of orbitals to describe Rydberg excitations. This strategy substantially reduces the computational cost by enabling the use of a larger set of virtual orbitals compared to an active space in which RAS2 spans both occupied and virtual orbitals, although convergence may be slower and less robust.

For XPS, the neutral ground state (GS) and the corresponding core-ionized GS are considered. Two SS-RASSCF calculations are performed, one for each state. Dynamic correlation is subsequently included via SS-RASPT2 to refine the ionization energies, whereas photoelectron intensities are estimated from Dyson norms evaluated at the RASSCF level, which quantify the overlap between the N- and (N$ - $1)-electron wavefunctions. \cite{pickup1977theory,mishra1980correlated,spanner2012dyson}
For XAS, the GS is computed at the SS-RASSCF level, while core-excited states are obtained with SA-RASSCF, followed by the evaluation of oscillator strengths. Dynamic correlation is then accounted for by (R)MS-CASPT2 on top of the SA-RASSCF manifold, improving excitation energies by capturing correlation outside the active space. \cite{schnorr2019tracing,nenov2019x,delcey2019efficient,northey2020ultrafast,segatta2020exploring,eckert2019t1,vacher2020origin}
Contributions from different conformers are finally weighted according to their Boltzmann populations following Ref. \citenum{myhre2019x}. 

Access to high-lying core states is enabled by the Highly Excited States (HEXS) projection technique. \cite{delcey2019efficient} In the absence of HEXS, the manifold would be dominated by lower-lying valence states. HEXS circumvents this issue by projecting out configuration state functions that fully occupy a chosen orbital subspace, effectively zeroing their CI coefficients. Applied to the subspace containing the target core orbitals, the method grants direct access to singly and multiply core-excited states (including those localized on different atomic centers) and allows computation of transition dipoles between valence, singly core-excited, and doubly core-excited states. \cite{schnorr2019tracing,nenov2019x,delcey2019efficient} 

The protocol presented above for gas-phase calculations is extended to the solution phase within the polarizable MCSCF/FQ embedding framework (see Fig.\ref{fig:protocol}). The computational workflow integrates MD simulations with the multiconfigurational electronic-structure RASSCF-RASPT2/FQ method, following the strategy developed by some of the present authors,~\cite{giovannini2020molecular,shepard1980multiconfigurational,gomez2022multiple} which implies:

\begin{enumerate}
    \item \textbf{System definition} – The system is partitioned into a QM region (the solute, treated with RASSCF/RASPT2) and the classical FQ region (the solvent, consisting of an appropriate number of atomistic water molecules in the present case.

    \item \textbf{Conformational sampling} – The configurational space of the solvated system is explored via nanosecond classical molecular dynamics (MD), ensuring statistically meaningful sampling of solute–solvent configurations (see Electronic Supplementary Information - ESI for details).

    \item \textbf{Snapshot extraction} – A set of uncorrelated configurations is selected from the MD trajectory. Each snapshot is truncated to a spherical droplet centered on the solute, with a cutoff radius chosen to retain most relevant solute–solvent interactions.

    \item \textbf{QM/FQ spectroscopy calculations} – For each snapshot, XPS and XAS spectra are computed using the RASSCF-RASPT2/FQ approach, applying the same RAS strategy as in the gas phase. HEXS is activated when direct access to core-excited configurations is required.

    \item \textbf{Spectral reconstruction} – Final spectra are obtained by averaging the results from all snapshots, followed by Gaussian convolution to account for inhomogeneous broadening. 
\end{enumerate}

\section{Computational Details}\label{sec:compdet}

MD simulations were performed using GROMACS 2020.4.\cite{abraham2015gromacs} Parameters for glycine were generated using ACPYPE-antechamber and the General Amber Force Field (gaff).\cite{acpypeSousa2012,kagami_acpype_2023,antechamberWang2006,gaffWang2004} The initial structure of the zwitterionic form of glycine was optimised using Gaussian 16 at the B3LYP/6-31+G(d) level of theory with water included as an implicit solvent using the cpcm method.\cite{g16} An electrostatic potential (ESP) grid was generated at the same level using the Merz-Singh-Kollman scheme and charges were produced through RESP fitting.\cite{MSKa,MSKb} The Boys localisation method was used to identify the centroid positions of oxygen lone pairs for inclusion of virtual sites.\cite{boys-RevModPhys.32.296} TIP5PE parameters were employed for water.\cite{tip5p-e10.1063/1.1652434} Glycine was then solvated in a 5.45~\AA~cubic box, comprising 5261 water molecules. The potential energy of the system was minimised using steepest descent. Periodic boundary conditions were used. A 0.9 nm cut-off was applied for short-range electrostatic and 
%Van de Waals 
van der Waals
interactions, while long range electrostatic interactions were treated using the particle-mesh Ewald (PME) method.\cite{darden1993particle} Using the canonical ensemble (NVT), the system was heated to 298.15 K and equilibrated for 1 ns using the velocity-rescaling method (coupling constant of 0.1 ps) and a time step of 2 fs. Constraints were applied to H-bonds using the LINCS algorithm.\cite{LINCSb,hess1997lincs} The density of the system was then equilibrated over 2 ns with an isothermal-isobaric ensemble (NPT) using the Parrinello-Rahman barostat (2 ps coupling constant) before a production run of 30 ns was subsequently performed from which 200 uncorrelated snapshots were extracted at regular 150 ps intervals. Spherical droplets of radius 17~\AA~centered on the solute were cut using MDanalysis.\cite{mdanalysis} H-bond analyses (radius= 3.5~\AA, angle 30$^{\circ}$) were performed using GROMACS while radical distribution functions between solute oxygen and nitrogen atoms and the solvent were calculated using MDanalysis.

XPS and XAS were computed for each molecular configuration using the RASSCF–RASPT2 approach in the gas phase and the RASSCF–RASPT2/FQ scheme in aqueous solution, employing a locally modified version of OpenMolcas. \cite{aquilante2020modern,li2023openmolcas}
Two FQ parametrizations were considered: FQ$^{a}$ from Ref.~\citenum{rick1994dynamical} and FQ$^{b}$ from Ref.~\citenum{giovannini2018effective}.
Gas-phase spectra were obtained as Boltzmann-weighted averages over the five lowest-energy glycine conformers (Fig. \ref{fig:SI_conformers}) using the structures and relative populations reported in Ref.~\citenum{myhre2019x}. 
For the solution phase, XPS and XAS spectra were computed over an initial ensemble of 200 snapshots. The number of successfully converged calculations depends on the edge and embedding model, as detailed for example in Fig \ref{fig:xas_conv}.

Active spaces were defined using the notation RAS1($a,b$)/RAS2($c,d$)/RAS3($e,f$). Here, $a$ denotes the number of core orbitals included in RAS1, with up to $b$ holes allowed; $c$ electrons are distributed over $d$ active orbitals in RAS2, which is treated at the full-CI level; and $f$ virtual orbitals are included in RAS3, with at most $e$ electrons permitted.
For XPS calculations, the active spaces RAS1(1,1)/RAS2(12,7)/RAS3(0,0) and RAS1(1,1)/RAS2(12,10)/RAS3(0,0) were employed. For XAS, the active spaces RAS1(1,1)/RAS2(12,6)/ RAS3(2,6) and RAS1(1,1)/RAS2(12,6)/RAS3(2,10) were used, including 
10 and 15 electronic states, respectively. 
All calculations were performed using the ANO-RCC-VTZP basis set.

To assess the role of solute–solvent polarization effects, additional non-polarizable QM/MM calculations were carried out using the electrostatic potential fitting (ESPF) scheme \cite{ferre2002approximate} with TIP3P point charges. \cite{jorgensen1981quantum}

In addition to the choice of the active space, the sensitivity of RASPT2 results to the IPEA shift \cite{ghigo2004modified} and to the frozen-core approximation \cite{finley1998multi} was examined for the main gas-phase conformer of glycine (see table \ref{tab:ipea_frozen} in the ESI). The IPEA shift was found to have a negligible impact on excitation energies, whereas the frozen-core treatment plays a critical role. Disabling the frozen-core approximation—thereby allowing explicit correlation of core orbitals at the RASPT2 level—is essential for an accurate description of core-level spectra. Notably, the data reported in table \ref{tab:ipea_frozen} reveals marked differences between the signals associated with the two oxygen atoms: their relative energy ordering is reversed when core orbitals are excluded from correlation, highlighting the importance of core–valence correlation effects.

\section{Results and Discussion}

\subsection{MD analysis and hydration pattern}

\begin{figure}[H]
    \centering
    \includegraphics[width=0.8\linewidth]{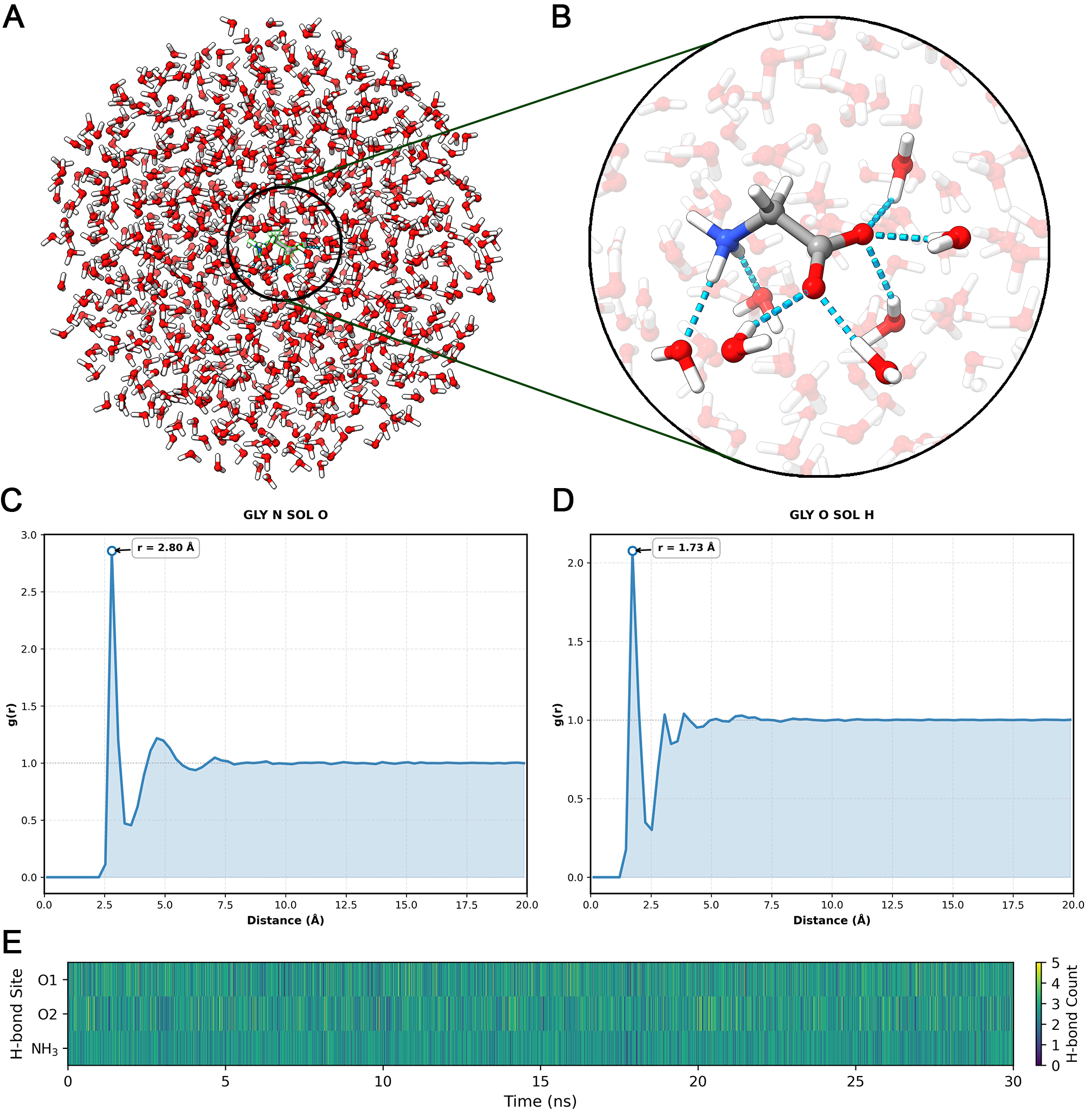}
    \caption{Analysis of hydration pattern during MD simulations of aqueous glycine. A) Representative solvated droplet at 0.6 ns with B) callout illustrating the 
\mbox{H-bond} pattern between waters and glycine. \mbox{H-bonds} shown with blue dashed lines. C) RDF between glycine N and water O. D) RDF between glycine carboxylate O's and water H's. E) Heatmap showing hydrogen bonding to glycine O1, O2, and NH$_3$ over time. Snapshots recorded ever 1 ps. }
    \label{fig:MDanalysis}
\end{figure}

In comparison to the neutral gas phase structures, the dominant zwitterionic form of glycine in aqueous solution is more symmetric and the main source of conformational variability results from the O-C-C-N dihedral angle (Figure \ref{fig:MDSI} in ESI). As shown in Figure \ref{fig:MDanalysis}, there is clear organization of the solvent around the two charged components: radial distribution functions (RDF) between the nitrogen and solvent oxygen, and carboxylate oxygens and solvent hydrogen, show significant peaks at 2.8~\AA~ and 1.73~\AA~, respectfully, with bulk properties returning at radii greater than 7.5~\AA. On average, 5.2 $\pm$ 0.9 H-bonds were present between the solvent and glycine carboxylate (2.6 per oxygen) while 2.6 $\pm$ 0.7 H-bonds were present between the glycine NH$_3$ and solvent. These H-bond patterns were more or less consistent throughout the duration of the 30 ns MD simulation (Figure \ref{fig:MDanalysis}E).

\subsection{XPS of glycine in gas-phase and aqueous solution}\label{sec:xps_results}

This section reports the computed XPS spectra of glycine at the nitrogen and carbon K-edges in both the gas phase and aqueous solution. Gas-phase spectra are obtained at the RASSCF–RASPT2 level, whereas condensed-phase spectra are computed using the RASSCF–RASPT2/FQ$^{(a,b)}$ models, following the protocol detailed in \cref{sec:xps_xas,sec:compdet}. Two active spaces, (12,7) and (12,10), are employed. The resulting spectra are compared with available experimental 
data~\cite{xps_exp_gas,xps_exp_water} and, for the aqueous phase, also with non-polarizable QM/ESPF calculations with TIP3P charges.

\begin{figure}[ht!]
    \centering
    \includegraphics[width=0.8\linewidth]{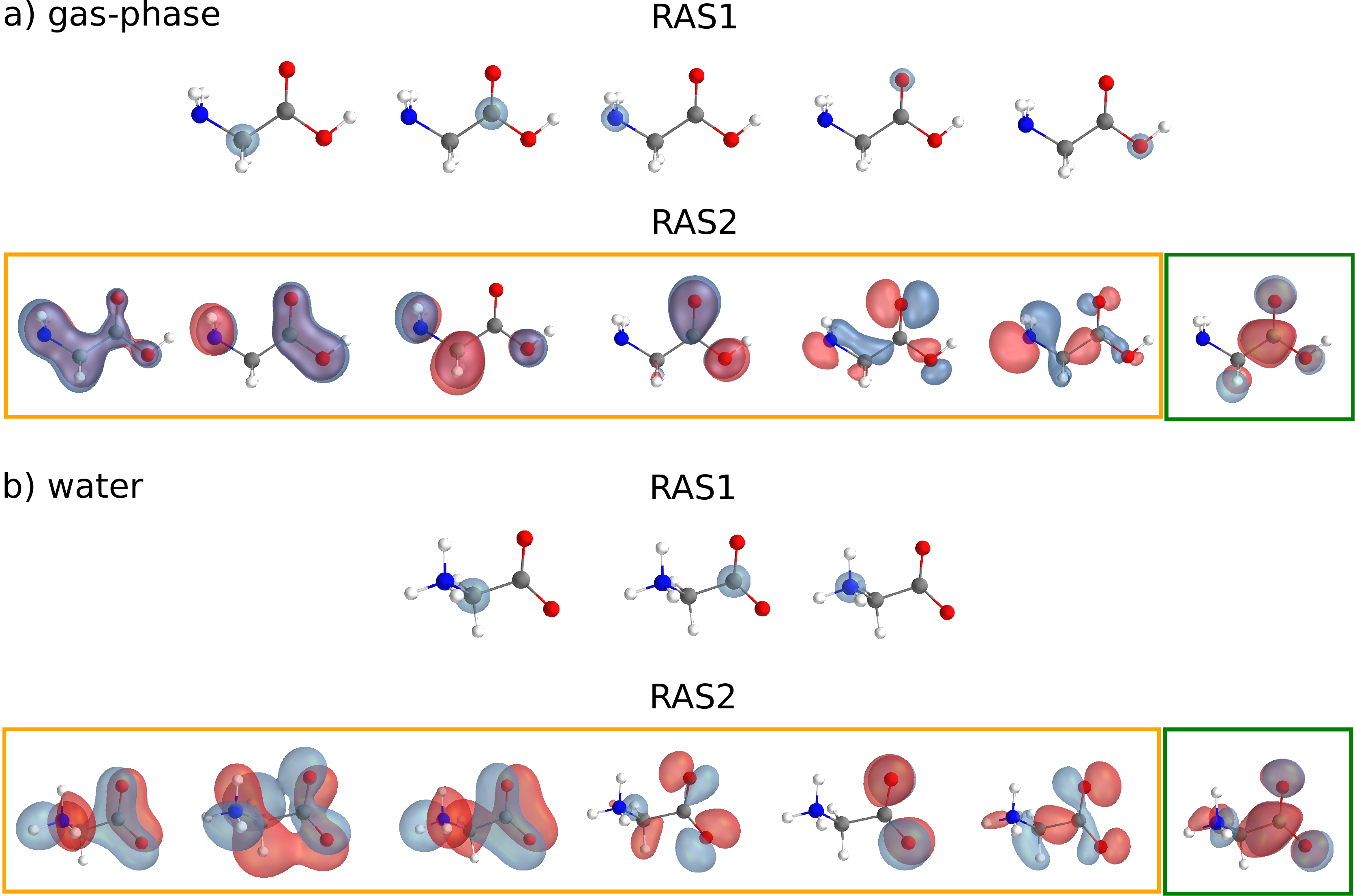}
    \caption{Hartree–Fock (HF) orbitals selected for the active space (12,7) in the RASSCF calculations of glycine: a) gas-phase (configuration I, as identified in Ref. \citenum{myhre2019x}); b) aqueous-phase zwitterionic configuration (obtained using the FQ$^a$ solvent model). The orbitals are classified into RAS1 and RAS2 subspaces: RAS1 includes the chemically relevant core orbital (one per calculation), while RAS2 contains selected occupied orbitals (outlined with a yellow rectangle) and virtual orbitals (outlined with a green rectangle).}
    \label{fig:HF_orb}
\end{figure}

In \cref{fig:HF_orb}, the HF orbitals included in the (12,7) active space are shown for both the gas-phase structure (\cref{fig:HF_orb}a, configuration I as identified in Ref.~\citenum{myhre2019x}) and for a representative solvated snapshot (\cref{fig:HF_orb}b), together with their assignment to the RAS1 and RAS2 subspaces. RAS1 contains the relevant core orbital (one per calculation). In the gas phase, RAS2 comprises six occupied orbitals—corresponding to the nitrogen and oxygen lone pairs (n$_{\text{N}}$, n$_{\text{O}}$) and the $\pi$ orbitals—together with one virtual $\pi^*$ orbital. In aqueous solution, glycine adopts a zwitterionic form, which leads to a markedly different molecular geometry with respect to the gas-phase configuration. As a consequence, the shape and spatial distribution of the orbitals also change. However, the chemical character of the orbitals selected for the active space—namely the n$_{\text{N}}$, n$_{\text{O}}$, and $\pi/\pi^*$ orbitals—remains the same (\cref{fig:HF_orb}b). This choice of occupied and virtual $\pi$ orbitals is consistent with the criteria adopted in Ref.~\citenum{montorsi2022soft}.
The larger active space (12,10) extends the (12,7) by including three additional Rydberg-type orbitals in the virtual manifold. % (as proposed in Ref. \citenum{muchova2007glycine}). 
The final RASSCF natural orbitals for each nucleus are reported in Figs. \ref{fig:orb_vacuo_127}, \ref{fig:orb_water_127}, \ref{fig:orb_vacuo_1210}, \ref{fig:orb_water_1210} 
in the ESI, where they are also grouped according to their assignment to the RAS1 and RAS2 subspaces, for both the gas phase and aqueous solution.

\begin{figure}[ht!]
    \centering
    \includegraphics[width=0.99\linewidth]{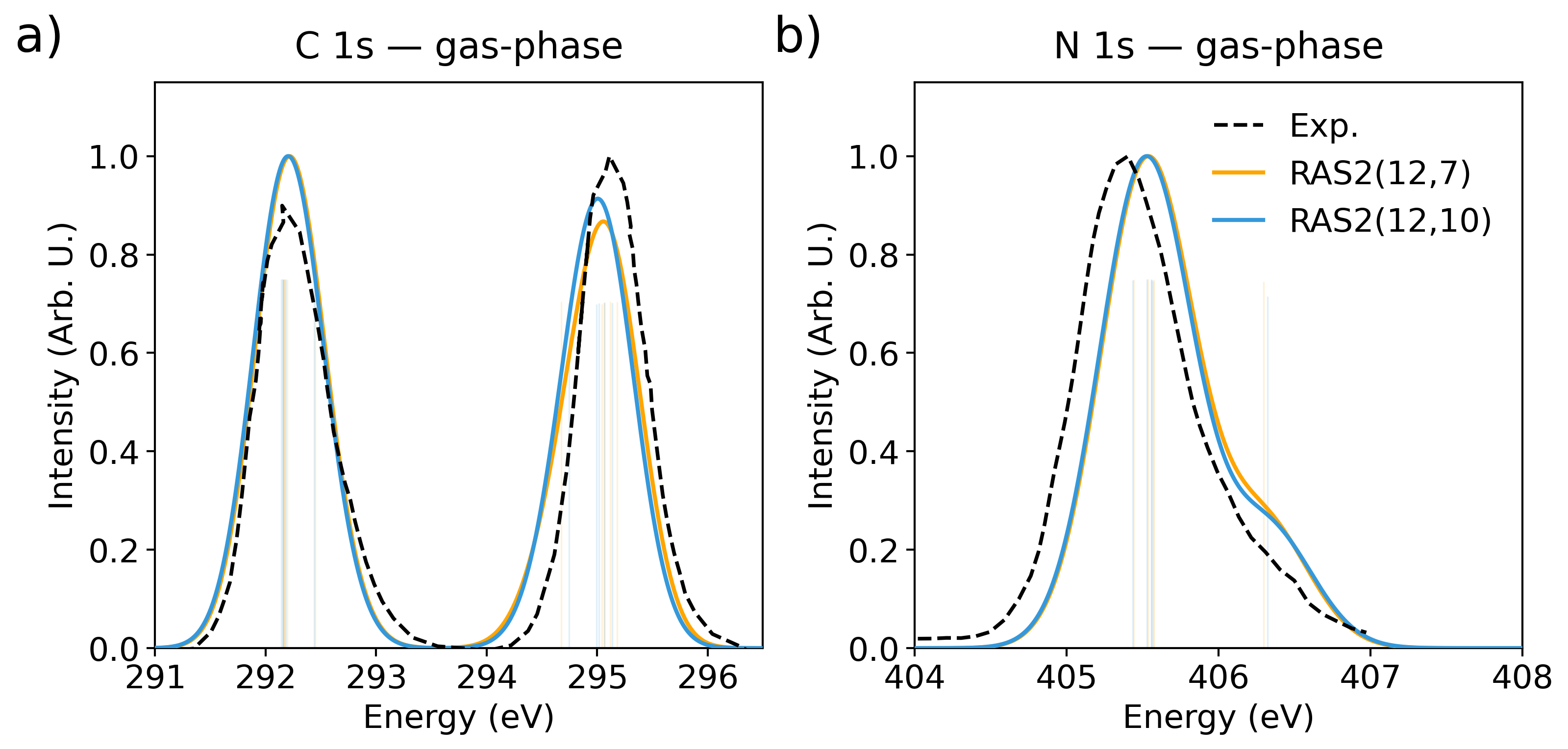}
    \caption{Computed C K-edge (a) and N K-edge (b) XPS spectra of glycine in the gas phase with two different active spaces, evaluated as a Boltzmann-weighted average over the five glycine conformers (see Ref. \citenum{myhre2019x} for the structures and their populations). All spectra were computed using the ANO-RCC-VTZP basis set and compared to experimental spectrum taken from Ref. \citenum{xps_exp_gas}.}
    \label{fig:XPS_gas}
\end{figure}

\begin{figure}[ht!]
    \centering
    \includegraphics[width=0.99\linewidth]{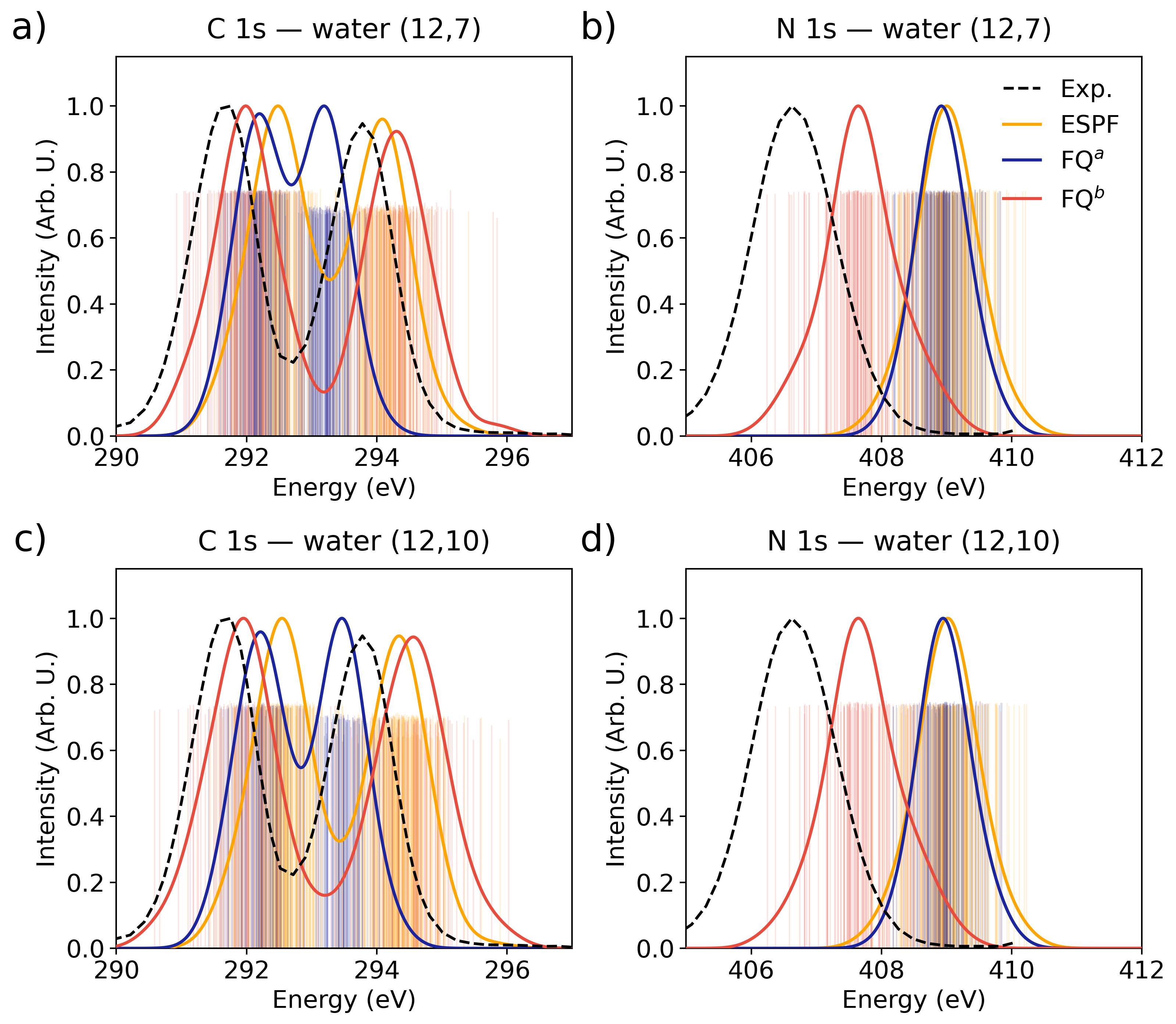}
    \caption{Aqueous-phase spectra obtained with RASSCF–RASPT2/ESPF and RASSCF–RASPT2/FQ$^{(a,b)}$, averaged over 200 MD snapshots. For all calculations, the corresponding stick spectra are also provided. All spectra were computed using the ANO-RCC-VTZP basis set. Experimental spectra take from Ref. \citenum{xps_exp_water} are also shown for comparison.}
    \label{fig:XPS_water}
\end{figure}

Computed XPS spectra in the gas phase and in aqueous solution are reported in \cref{fig:XPS_gas} and \cref{fig:XPS_water}, respectively. 
The left and right panels of \cref{fig:XPS_gas} show the gas-phase spectra computed with the (12,7) and (12,10) active spaces, respectively, obtained as a Boltzmann-weighted average over the five glycine conformers (shown in Fig \ref{fig:SI_conformers} in ESI, see Ref. \citenum{myhre2019x} for computation details and their populations).
In solution, the spectra shown on the top ((12,7) active space) and bottom ((12,10) active space) panels of \cref{fig:XPS_water} were obtained by averaging over 200 uncorrelated snapshots extracted from the MD trajectory, using the QM/ESPF and the QM/FQ$^{(a,b)}$ models. For each snapshot, the computed photoionization energies and Dyson norms are also reported 
in~\cref{fig:XPS_water} as stick spectra. 
The spread across snapshots reflects the combined influence of solvent dynamics and solute conformational fluctuations, which together produce inhomogeneous broadening. Convergence with respect to the number of snapshots is verified (see Figs. \ref{fig:c_conv_127}, \ref{fig:n_conv_127}, \ref{fig:c_conv_1210}, and \ref{fig:n_conv_1210} in the ESI).
 system. 
The spectra become essentially stable upon inclusion of approximately
100--150 successfully converged configurations, depending on the edge,
active space, and FQ parametrization 
(see section \ref{sec:XPS_conv}).

The gas-phase C K-edge spectrum of glycine (see \cref{fig:XPS_gas}a) exhibits two distinct peaks at 292.2 eV and 295.0 eV for both the (12,7) and (12,10) active spaces, assigned to ionization from the C$_{\textsc{H$_2$}}$ and C$_{\textsc{O}}$ carbon atoms, respectively, with an energy separation of 2.8 eV. The figure also reports the contributions from individual conformers: the two principal bands originate predominantly from conformer 1, which accounts for 73.2\% of the Boltzmann population, while minor contributions from conformer 2 (13.2\%) lead to a slight inhomogeneous broadening of the spectrum (see Ref.~\citenum{myhre2019x} for the Boltzmann populations). 
The experimental spectrum \cite{xps_exp_gas} exhibits two peaks at 292.3 and 295.2 eV, separated by 2.9 eV. The computed spectra thus reproduce these values with remarkable accuracy. The main discrepancy concerns the relative intensities: in the experimental spectrum the higher-energy peak is the most intense, whereas in the computed spectra it is the weakest. This limitation is slightly alleviated by the (12,10) active space, which increases the intensity of the second peak compared to the (12,7) case.

In solution, for the (12,7) active space (see \cref{fig:XPS_water}a), the peaks associated with the C$_\textsc{H$_2$}$ and C$_\textsc{O}$ atoms are located at 292.5 eV and 294.1 eV for RASPT2/ESPF, 292.2 eV and 293.2 eV in the RASSCF-RASPT2/FQ$^{a}$ model, and at 292.0 eV and 294.3 eV in RASSCF-RASPT2/FQ$^{b}$, respectively. The resulting separations are 1.6 eV for RASPT2/ESPF, 1.0 eV for RASSCF-RASPT2/FQ$^{a}$ and 2.3 eV for RASSCF-RASPT2/FQ$^{b}$. With the (12,10) active space (see \cref{fig:XPS_water}c), the corresponding peaks appear at 292.5 eV and 294.3 eV for RASPT2/ESPF, 292.2 eV and 293.5 eV for RASSCF-RASPT2/FQ$^a$, and at 291.9 eV and 294.6 eV for FQ$^b$, yielding separations of 1.8 eV, 1.3 eV and 2.6 eV, respectively. Particular attention should be paid to the relative intensities: with the FQ$^a$ parameters, the higher-energy peak is the most intense, in contrast to the results obtained with FQ$^b$. This trend is consistent across both active spaces. 
The experimental spectrum \cite{xps_exp_water} shows two peaks at 291.8 and 293.8 eV, with a separation of 2.0 eV. Both RASSCF-RASPT2/FQ$^a$ and RASSCF-RASPT2/FQ$^b$ reproduce the absolute position of the first peak with good accuracy, with the (12,10) active space being in slightly better agreement. Larger deviations, however, are found for the second peak, associated with the C$_\textsc{O}$ atom, and the corresponding peak separation. The FQ$^b$ (12,7) values are the closest to the experimental separation of 2.0 eV, as highlighted in \cref{fig:XPS_water}(a,c). Furthermore, the RASPT2/FQ$^b$ spectra also better reproduce the relative intensities and the experimental band broadening, whereas RASPT2/FQ$^a$ with both active spaces fails to capture the correct spectral profile. Overall, RASPT2/FQ$^b$ emerges as the most reliable parametrization for describing the C K-edge 
XPS
in solution.

The gas-phase N K-edge XPS spectrum of glycine (\cref{fig:XPS_gas}b) exhibits a main peak at 405.5 eV, and a secondary peak at about 406.3 eV, which is consistently reproduced by both active spaces. The computed spectra are nearly identical regardless of the active space used. The conformer-resolved contributions indicate that the main band is dominated by conformer 1 (73.2\% Boltzmann population), while the secondary peak is mainly due to conformer 2 (13.2\%).
The gas-phase experimental spectrum displays a single prominent peak located at 405 eV. The theoretical spectra accurately reproduce both the position and its shape. 

In solution, the corresponding peaks are located at 408.9 eV (RASPT2/ESPF, RASPT2/FQ$^{a}$) and 407.7 eV (RASPT2/FQ$^{b}$) for both active spaces (\cref{fig:XPS_water}d). The (12,10) active space leads to 409.0 eV for both RASPT2/ESPF, RASPT2/FQ$^{a}$ and 407.7 eV for RASPT2/FQ$^{b}$, and yields broader bands than (12,7). Differences between the two FQ parametrizations are more pronounced: peak positions differ by more than 1~eV, and the FQ$^{b}$ spectrum exhibits noticeably larger inhomogeneous broadening.
The experimental peak is observed at 406.6 eV and, again, RASPT2/FQ$^b$ is the method that provides the best results compared with the experimental data, also taking into account the broadening of the band.

The gas-phase spectra are in very good agreement with the experimental data, both in terms of peak positions and relative intensities, confirming that the chosen active spaces and the RASSCF–RASPT2 protocol provide a reliable description of the core-ionized states. When moving to solution, however, the situation becomes more delicate. The computed spectral features—peak positions, relative intensities, and band widths—exhibit a marked dependence on the parametrization of the FQ model. 
In all cases, the theoretical spectra are shifted to higher energies relative to experiment, indicating that the solvent-induced stabilization of the cationic state is underestimated compared with the stabilization of the ground state at the same level of theory.  Among the two FQ parametrizations tested, FQ$^{b}$ yields a noticeably improved agreement with experiment, likely due to its more balanced treatment of solute–solvent electrostatics and polarization effects,\cite{giovannini2019effective} where FQ$^{a}$ was specifically optimized to reproduce the properties of bulk water instead.

\subsection{XAS of glycine in gas-phase and aqueous solution}

Nitrogen and carbon K-edge XAS spectra of glycine in gas-phase spectra are computed at the RASSCF–RASPT2 level, while spectra in solution are obtained using the RASSCF–RASPT2/FQ$^{(b)}$ and RASSCF–RASPT2/ESPF (with TIP3P point charges) schemes, following the computational protocol described in \cref{sec:xps_xas,sec:compdet}. 
The calculated spectra are validated through comparison with available experimental data. \cite{plekan2007x,messer2005ph}

The number of roots and the active spaces are selected to describe all relevant core-to-valence excitations; accordingly, a relatively large set of virtual orbitals is included. The HF orbitals placed in the active space comprise the nitrogen and oxygen lone pairs (n$_{\text{N}}$, n$_{\text{O}}$), the occupied $\pi$ orbitals, one $\pi^*$ orbital, and a variable number (either 6 or 10) of $\sigma^*$ and/or Rydberg-type virtuals. Specifically, two active-space are used: RAS1(1,1)/RAS2(12,6)/RAS3(2,6) and RAS1(1,1)/RAS2(12,6)/RAS3(2,10). Here, the notation RAS1($a,b$)/RAS2($c,d$)/RAS3($e,f$) denotes $a$ core orbitals in RAS1 allowing up to $b$ holes, $c$ electrons distributed over $d$ active orbitals in RAS2, and $f$ virtual orbitals in RAS3 with at most $e$ electrons. 
As discussed for XPS, glycine adopts a zwitterionic structure in aqueous solution, which markedly alters the spatial distribution and energies of the molecular orbitals. However, the chemical character of the frontier orbitals defining the active space remains unchanged. The orbitals comprising these active spaces are shown in figs. \ref{sec:SI_xas_c} and \ref{sec:SI_xas_n} in ESI.
Seeing how the larger RAS3(2,10) space introduces an extra transition in the N 
K-edge spectra, lying in a forbidden continuum energy region, this section focuses exclusively on spectra from the smaller RAS3(2,6) active space. 

\begin{figure}[b!]
    \centering
    \includegraphics[width=0.9\linewidth]{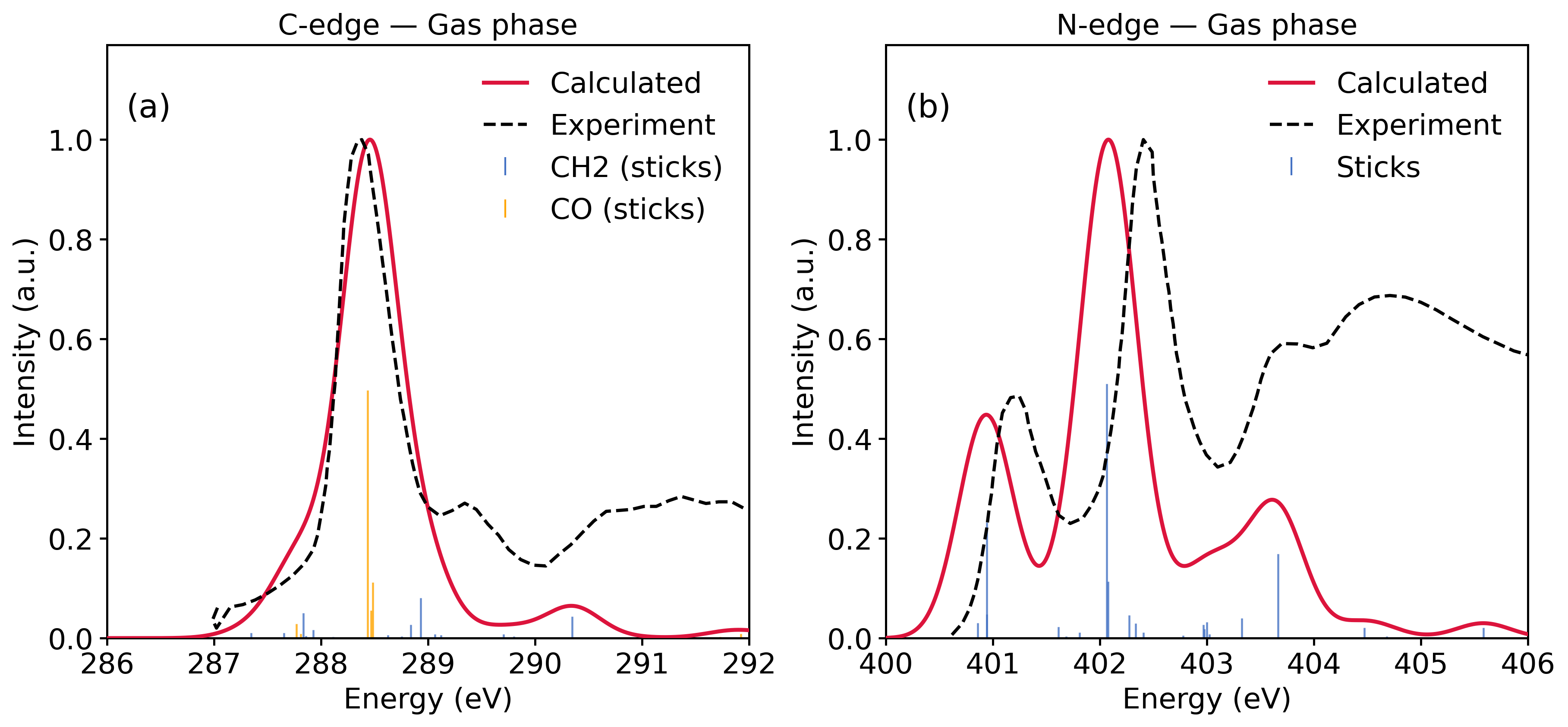}
    \caption{RASSCF-RASPT2 XAS spectra in the gas-phase calculated with the RAS1(1,1)/RAS2(12,6)/RAS3(2,6) active space averaged over 10 states, evaluated as a Boltzmann-weighted average over the five glycine conformers (see ESI and Ref. \citenum{myhre2019x} for structures and populations). All spectra were computed using the ANO-RCC-VTZP basis set, and compared against experimental values taken from Refs. \citenum{plekan2007x} and \citenum{messer2005ph}.}
    \label{fig:XAS_gas}
\end{figure}
\begin{figure}[t!]
    \centering
    \includegraphics[width=0.9\linewidth]{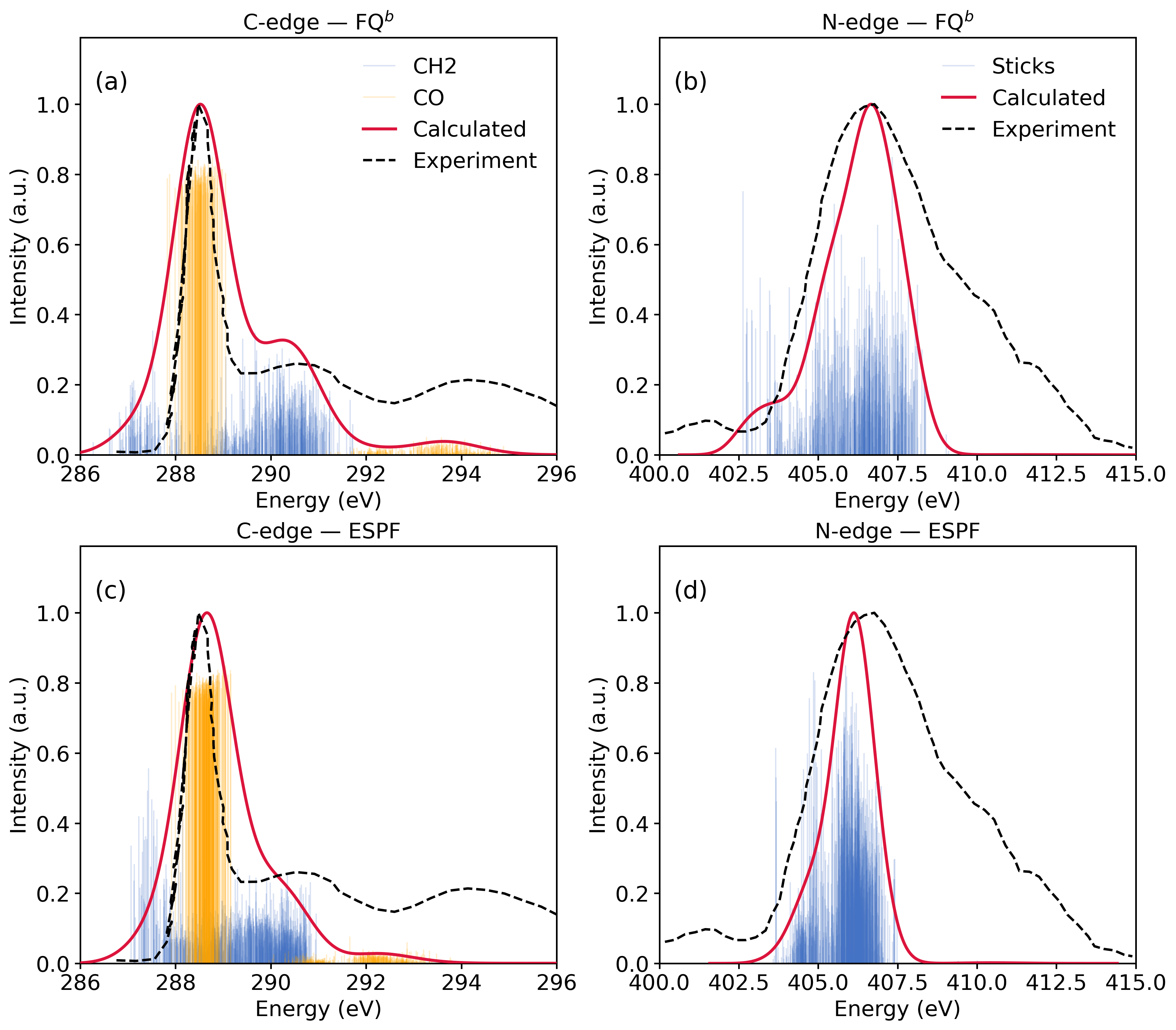}
    \caption{Aqueous-phase spectra obtained with RASSCF–RASPT2/ESPF and RASSCF–RASPT2/FQ$^{(a,b)}$, using the RAS1(1,1)/RAS2(12,6)/RAS3(2,6) active space averaged over 10 states. All calculations were averaged over 200 MD snapshots, and the corresponding stick spectra are provided in the image. All spectra were computed using the ANO-RCC-VTZP basis set, and compared against experimental values taken from Refs. \citenum{plekan2007x} and \citenum{messer2005ph}.}
    \label{fig:XAS_solv}
\end{figure}

The stick and the convoluted RASSCF-RASPT2 spectra for gas-phase and solvated glycine are collected in 
\cref{fig:XAS_gas} and \cref{fig:XAS_solv} together with the corresponding experimental data. \cite{plekan2007x,messer2005ph} For the gas phase (see \cref{fig:XAS_gas}), panels (a),(b) show the unshifted spectra at the C and N K-edges.

Gas-phase spectra are obtained as Boltzmann-weighted averages over the five most populated conformers identified in Ref. \citenum{myhre2019x}. For the condensed phase 
(see \cref{fig:XAS_solv}), 
spectra are computed over an ensemble of 200 snapshots and subsequently averaged. As in the XPS case, each snapshot yields a distinct spectrum, contributing to solvent-induced inhomogeneous broadening. While some difficulties are encountered in converging many of these snapshots, the overall convergence of the spectra with respect to the number of snapshots is still achieved after $\sim$100 frames; see ESI, Fig. \ref{fig:xas_conv}. \\

Focusing on the gas-phase C K-edge XAS 
(\cref{fig:XAS_gas}a),
the spectrum is characterized by a dominant band centered at 288.46 eV, followed by a weaker feature at 290.34 eV.
The stick spectra are further resolved into site-specific contributions from the carbonyl carbon (CO, yellow) and the methylene carbon 
(CH2, blue), with conformer-dependent intensities weighted by their Boltzmann populations. The intense low-energy peak is assigned to the C$_\textsc{CO}$ 1s $\rightarrow$ lowest RAS3 excitation (orbital 1; see fig.~\ref{fig:orbitals_co_gas} in the ESI). At higher energies, additional structure mainly arises from C$_{\textsc{CH}_2}$ 1s excitations into more diffuse RAS3 virtual orbitals—most prominently orbitals 1, 4, and 6 (fig.~\ref{fig:orbitals_ch2_gas}, ESI)—giving rise to weaker bands around 289.14, 290.06, and 291.68 eV. Conformer-resolved contributions are discussed in fig. \ref{fig:conform_C} in ESI: the three main bands are dominated by conformer 1 (73.2\% Boltzmann weight), while conformers 2 (13.2\%) and 4 (8.2\%) provide smaller contributions that slightly broaden the main peak through inhomogeneous effects.

At the C K-edge of solvated glycine, the ensemble-averaged RASSCF/RASPT2 spectrum, obtained with the small active space %(\cref{fig:XAS_gas_solv}(i)) 
(\cref{fig:XAS_solv}(c)) 
and ESPF solvent, shows a main peak at 288.52 eV and a weaker feature at 290.32 eV .

When using FQ$^{(b)}$ %(\cref{fig:XAS_gas_solv}(e)) 
(\cref{fig:XAS_solv}(a))
the two main features are located at 288.52 eV and 290.46 eV. The lowest-energy transition has predominant C$_\textsc{CO}$ 1s $\rightarrow$ $\pi^*$ character, whereas the higher-energy band shows mixed C$_\textsc{CH$_2$}$ 1s $\rightarrow$ $\sigma^*$ and Rydberg contributions.

Relative to the gas phase, solvation leads to marked inhomogeneous broadening and to a modest shift of about 0.059 eV in the main band. 

Conversely, the gas-phase N K-edge XANES spectrum of glycine (computed with the RAS3 = (2,6) active space, \cref{fig:XAS_gas}(b)) 
displays three well-separated bands at 400.94, 402.08, and 403.63 eV. The first feature is assigned to the N 1s $\rightarrow$ RAS3 orbital 5 transition, while the second—and most intense—band corresponds to excitation into RAS3 orbital 6 (see \ref{fig:orbitals_n_gas} in the ESI). The weaker high-energy peak at 403.61 eV is instead attributed to a transition into RAS3 orbital 1.

As in the C K-edge case, conformer-resolved stick spectra (fig. \ref{fig:conform_N} in ESI) indicate that the dominant contributions to the main bands arise from conformer 1 (73.2\% Boltzmann weight). Smaller contributions from conformers 2 (13.2\%) and 4 (8.2\%) slightly broaden the strongest peak through inhomogeneous effects.

At the N K-edge of solvated glycine, the ensemble-averaged RASSCF/RASPT2 spectra obtained with ESPF 
(\cref{fig:XAS_solv}(d)) 
show a main broad peak at 406.08 eV and a weaker feature at 404.67 eV. When using FQ$^{(b)}$
(\cref{fig:XAS_solv}(b)), 
a dominant band appears at 406.66 eV with a second peak at 403.50 eV. The lowest-energy transition can be assigned to the N 1s $\rightarrow$ $\pi^*$ transition (RAS3 orbital 1), whereas the main band is generated by various different ones. 

Comparing the theoretical and experimental data, the computed C K-edge spectrum accurately reproduces both the shape and peak position of the experimental one.
Both the overall band broadening and the relative peak separations are well captured, resulting in very good agreement between theory and experiment.

When compared with experimental data, the theoretical N K-edge spectrum slightly overestimates the peak energies. Regarding band broadening and peak separation, the agreement between theory and experiment is quite satisfactory.

\section{Summary and future perspectives}
\label{sec:summary}

In this work, we have presented the first application of the 
RASSCF--RASPT2/FQ implementation of the MCSCF/FQ polarizable embedding 
framework to the simulation of X-ray photoelectron and X-ray absorption 
spectra in aqueous solution. The approach combines a multireference treatment 
of core-ionized and core-excited states with an explicit, atomistic description 
of the solvent and configurational sampling from classical molecular dynamics. 
Glycine was selected as an initial benchmark because it provides a stringent 
test of the methodology: its neutral gas-phase conformers are replaced in 
water by a strongly hydrated zwitterionic structure, whose charged functional 
groups generate a structured and anisotropic solvation environment.

The gas-phase carbon and nitrogen K-edge spectra are quite well described by the RASSCF--RASPT2 protocol, confirming the suitability of the adopted active spaces and correlation treatment for the core-ionized and core-excited states 
of glycine. 

The aqueous spectra represent a more demanding test. Ensemble averaging over molecular-dynamics configurations is essential to recover the 
inhomogeneous broadening arising from fluctuations of both the solute and its 
hydration shell. For XPS, peak positions, relative intensities, and spectral 
widths show a pronounced dependence on the description of the environment. 
Among the models considered here, the FQ$^{b}$ parametrization provides the 
most balanced description of the carbon and nitrogen K-edges and improves the 
agreement with experiment relative to both FQ$^{a}$ and fixed-charge 
electrostatic embedding. Residual overestimations of the aqueous ionization 
energies nevertheless indicate that the relative stabilization of the neutral 
and core-ionized states is not yet fully captured.

For XAS, at the carbon K-edge, the FQ$^b$ model  reproduces the position, band structure and broadening of the experimental signal close to the absolute energy scale; similar results are achieved using the ESPF model as well. 
At the nitrogen K-edge, the main peak envelope is nicely reproduced by the FQ$^b$ model, albeit a slight underestimation of overall broadening and peak separation. Once again, the ESPF model also delivers a reasonable spectrum, but performs slightly worse than FQ$^b$ on all counts (absolute peak energy, broadening and peak separation). 
Taken together, these results show that a reliable simulation of core-level spectra in solution requires the simultaneous treatment of electronic correlation, the atomistic hydration structure, mutual solute--solvent  polarization, and statistically meaningful configurational sampling. They also show that the accuracy of a polarizable embedding model depends critically on 
the parametrization of its environmental response.

Although glycine constitutes a demanding first application, a definitive 
assessment of the accuracy and transferability of the method will require 
systematic tests on a broader range of molecular systems. Future benchmarks 
should include neutral and ionic solutes, different hydrogen-bonding patterns, 
and chemically distinct core sites, thereby allowing the dependence of the 
results on molecular charge, local hydration, and elemental edge to be 
separately investigated.

Further improvements may also arise from extending the present 
electrostatic-polarization coupling to include non-electrostatic contributions 
to the short-range solute--solvent interaction. In particular, Pauli repulsion \cite{giovannini2017general,giovannini2019quantum} and dispersion may affect the 
differential stabilization of the ground, core-ionized, and core-excited states. 
Their inclusion could reduce the remaining systematic energy errors and make 
the model less dependent on error compensation within a specific FQ 
parametrization.

A complementary development consists in enlarging the quantum-mechanical 
region to include selected water molecules from the first hydration shell. 
Such a treatment would permit a more explicit description of localized 
hydrogen bonds, solute--solvent orbital overlap, and possible charge-transfer 
contributions that cannot be represented by a purely classical environment. 
Preliminary calculations along these lines are currently being performed for 
glycine. Orbital-localization and active-space selection strategies are being 
explored to incorporate the most relevant solvent orbitals while preserving a 
well-defined and computationally tractable description of the targeted 
core-ionized and core-excited states.

The present study therefore, represents a proof of concept rather than a final 
benchmark of the RASSCF--RASPT2/FQ methodology. Nevertheless, the encouraging 
agreement obtained for both XPS and XAS demonstrates the potential of 
multireference polarizable embedding as a general framework for core-level 
spectroscopy in complex environments. Its extension to larger solvated 
molecules and, ultimately, to transient X-ray spectra of electronically excited 
systems constitutes a natural direction for future work.

\begin{acknowledgement}
This work was supported by the European Union's Horizon Europe Research and Innovation Programme under Grant Agreement  101169312. The authors gratefully acknowledge the Center for High-Performance Computing (CHPC) at the Scuola Normale Superiore for providing access to the computational infrastructure used in this work.
\end{acknowledgement}

% =================================================================
\section*{Author Declarations}
% =================================================================

\subsection*{Conflict of Interest}

The authors have no conflicts to disclose.

\subsection*{Author Contributions}

Chiara Sepali: data curation, formal analysis, investigation, software, writing - original draft.

Pietro Dal Cin: data curation; formal analysis; visualization, writing - original draft, writing - review and editing.

Alicia M. Kirk: data curation; formal analysis; visualization, writing - original draft, writing - review and editing.

Sonia Coriani: Conceptualization, Methodology, Supervision, Writing--review and editing.

Chiara Cappelli: Conceptualization, Methodology, Supervision, Project
administration, Funding acquisition, Writing--review and editing.

\section*{Supplementary Material}

Molecular structures and configurational sampling; assessment of the RASPT2 computational parameters; additional data for XPS and XAS of glycine in gas-phase and aqueous solution;%

\input{bbl}

\appendix
\input{SI}

\end{document}

%% file: bbl.tex
%aipnum4-2.bst 2019-01-14 (MD) hand-edited version of apsrev4-1.bst
%Control: key (0)
%Control: author (8) initials jnrlst
%Control: editor formatted (1) identically to author
%Control: production of article title (-1) disabled
%Control: page (0) single
%Control: year (1) truncated
%Control: production of eprint (0) enabled
%

%% file: SI.tex
\setkeys{acs}{articletitle = true}
\renewcommand{\thefigure}{S\arabic{figure}}
\setcounter{figure}{0}

\setcounter{table}{0}
\renewcommand{\thetable}{S\arabic{table}}
\setcounter{figure}{0}
\renewcommand{\thefigure}{S\arabic{figure}}
\setcounter{equation}{0}
\renewcommand{\theequation}{S\arabic{equation}}
\setcounter{page}{1}
\renewcommand{\thepage}{S\arabic{page}}
\setcounter{section}{0}
\renewcommand{\thesection}{S\arabic{section}}

\newpage

\section{Molecular structures and configurational sampling}

\begin{figure}[ht!]
    \centering
    \includegraphics[width=0.75\linewidth]{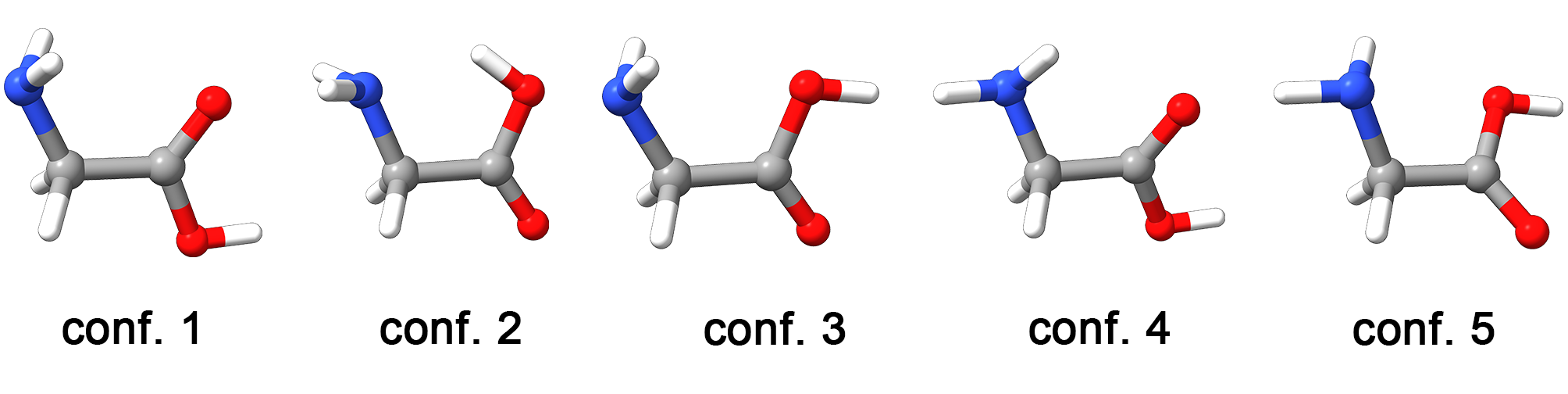}
    \caption{The five conformers of glycine in the gas-phase, taken from Ref. \citenum{myhre2019x}}
    \label{fig:SI_conformers}
\end{figure}

\subsection{MD simulations}

\begin{figure}[ht!]
    \centering
    \includegraphics[width=1\linewidth]{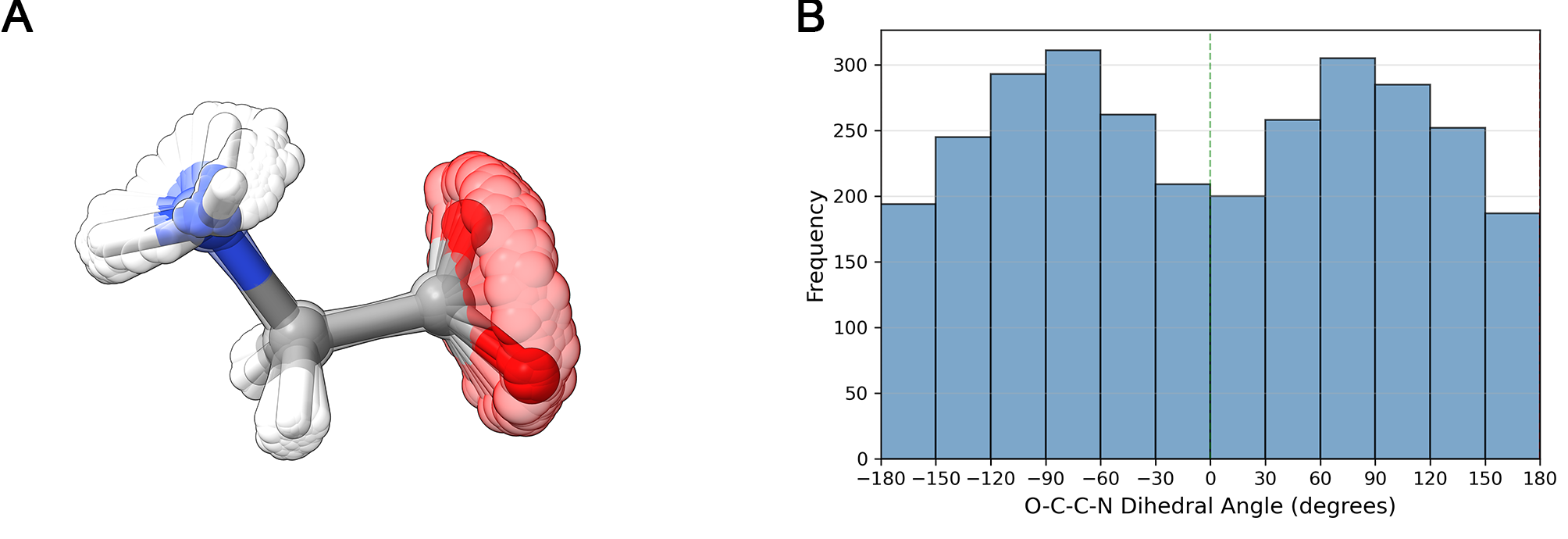}
    \caption{A) Overlap of 200 MD snapshots illustrating the conformational space explored by glycine. B) Histogram showing the O-C-C-N dihedral during the MD simulations.}
    \label{fig:MDSI}
\end{figure}

\newpage

\section{Assessment of the RASPT2 computational parameters}

\begin{table}[]
    \centering
    \begin{tabular}{|c|c|c|c|c|c|}
    \hline
     CASPT2 parameters & C$_{H_2}$ & C$_O$ & N & O$_C$ & $O_H$ \\
    Ipea = 0.0, Frozen=0 & 292.2 & 295.1 & 405.5 & 539.2 & 541.0  \\
    Ipea = 0.25, Frozen=0 & 292.2 & 295.1 & 405.6 & 539.2 & 541.0  \\
    Ipea = 0.0, No Frozen & 292.8 & 295.0 & 405.3 & 542.0 & 540.9  \\
    \hline
    \end{tabular}
    \caption{Core-ionization energies, in eV, calculated for the most populated
gas-phase conformer of glycine using different IPEA shifts and frozen-core
settings. The labels CH$_2$, CO, N, O$_{\mathrm{C}}$, and O$_{\mathrm{H}}$ identify the corresponding core-ionized atomic sites. Frozen = 0: frozen-core approximation enabled. No Frozen: Frozen-core approximation disabled.}
    \label{tab:ipea_frozen}
\end{table}

\section{XPS of glycine in gas-phase and aqueous solution}\label{sec:SI_xps}

\subsection{Molecular orbitals}

\begin{figure}[ht!]
    \centering
    \includegraphics[width=0.75\linewidth]{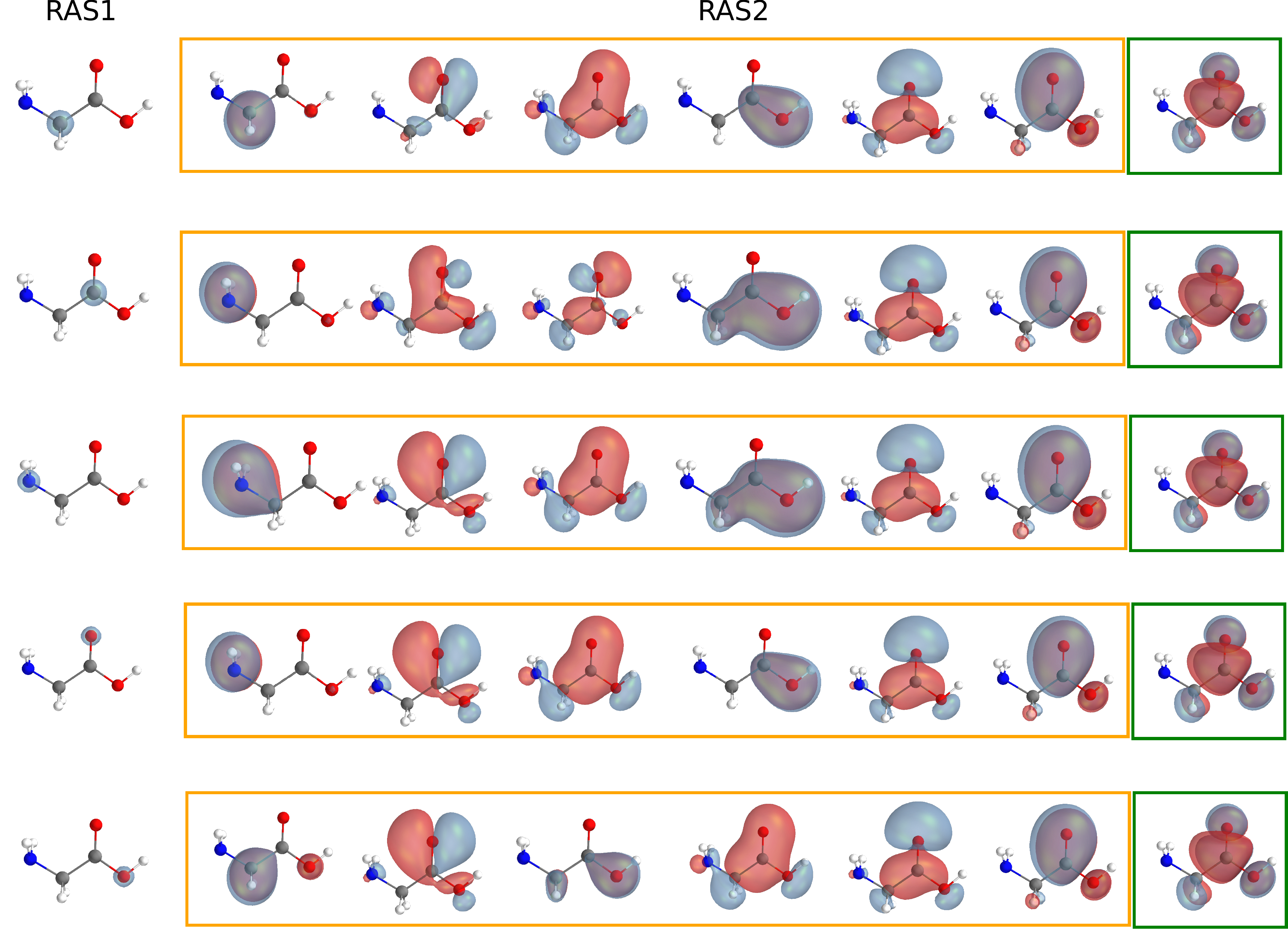}
    \caption{CASSCF natural orbitals corresponding to the active space (12,7) of glycine in the gas phase, shown separately for each nucleus: methylene carbon (C$_{H_2}$), carboxylic carbon (C$_O$), nitrogen (N), carbonyl oxygen (O$_C$), and hydroxyl oxygen (O$_H$). The orbitals are classified into RAS1 and RAS2 subspaces: RAS1 contains the chemically relevant core orbital (one per calculation), while RAS2 includes selected occupied orbitals (circled by a yellow rectangle) and virtual orbitals (circled by a green rectangle).}
    \label{fig:orb_vacuo_127}
\end{figure} 

\begin{figure}[ht!]
    \centering
    \includegraphics[width=0.75\linewidth]{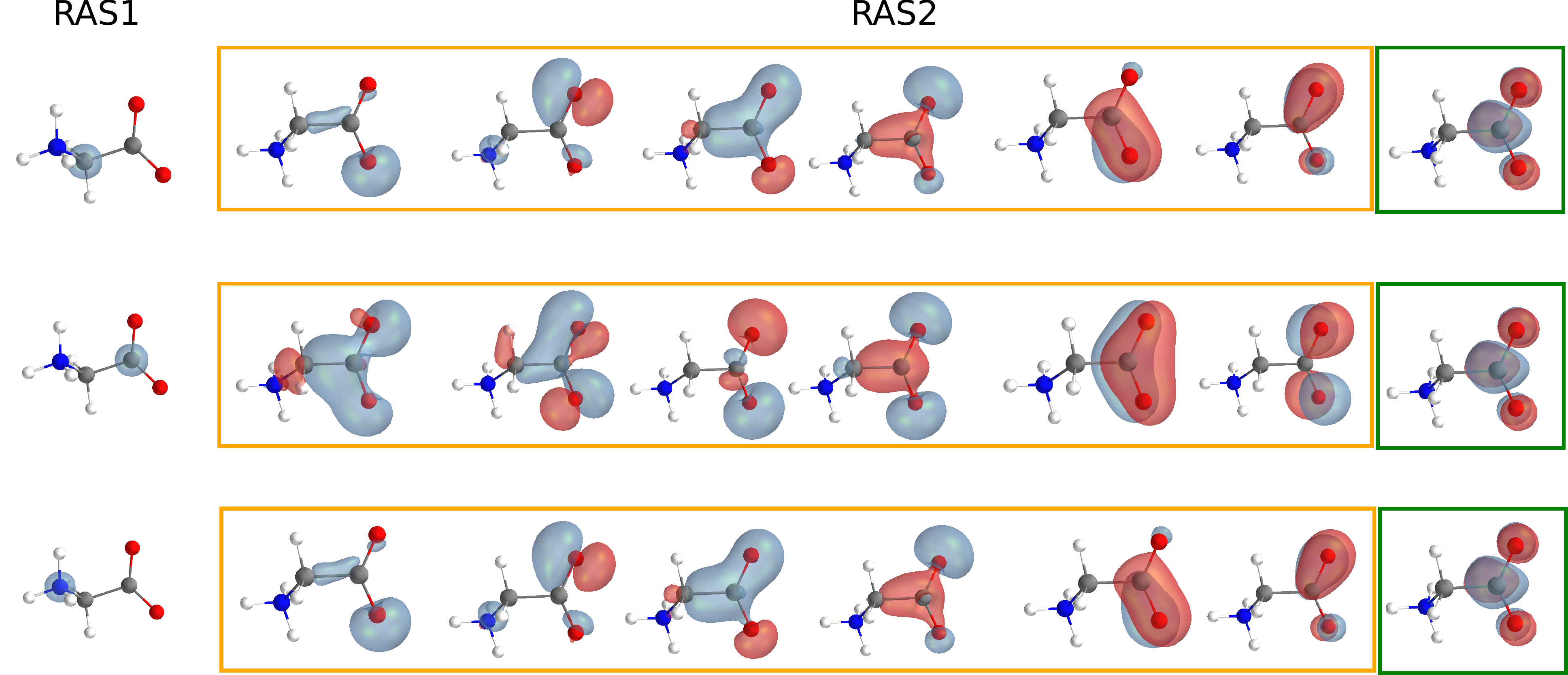}
    \caption{CASSCF natural orbitals corresponding to the active space (12,7) of a representative snapshot of glycine in aqueous solution, shown separately for each nucleus: methylene carbon (C$_{H_2}$), carboxylic carbon (C$_O$), nitrogen (N). The orbitals are classified as above.}
    \label{fig:orb_water_127}
\end{figure}

\begin{figure}[ht!]
    \centering
    \includegraphics[width=0.9\linewidth]{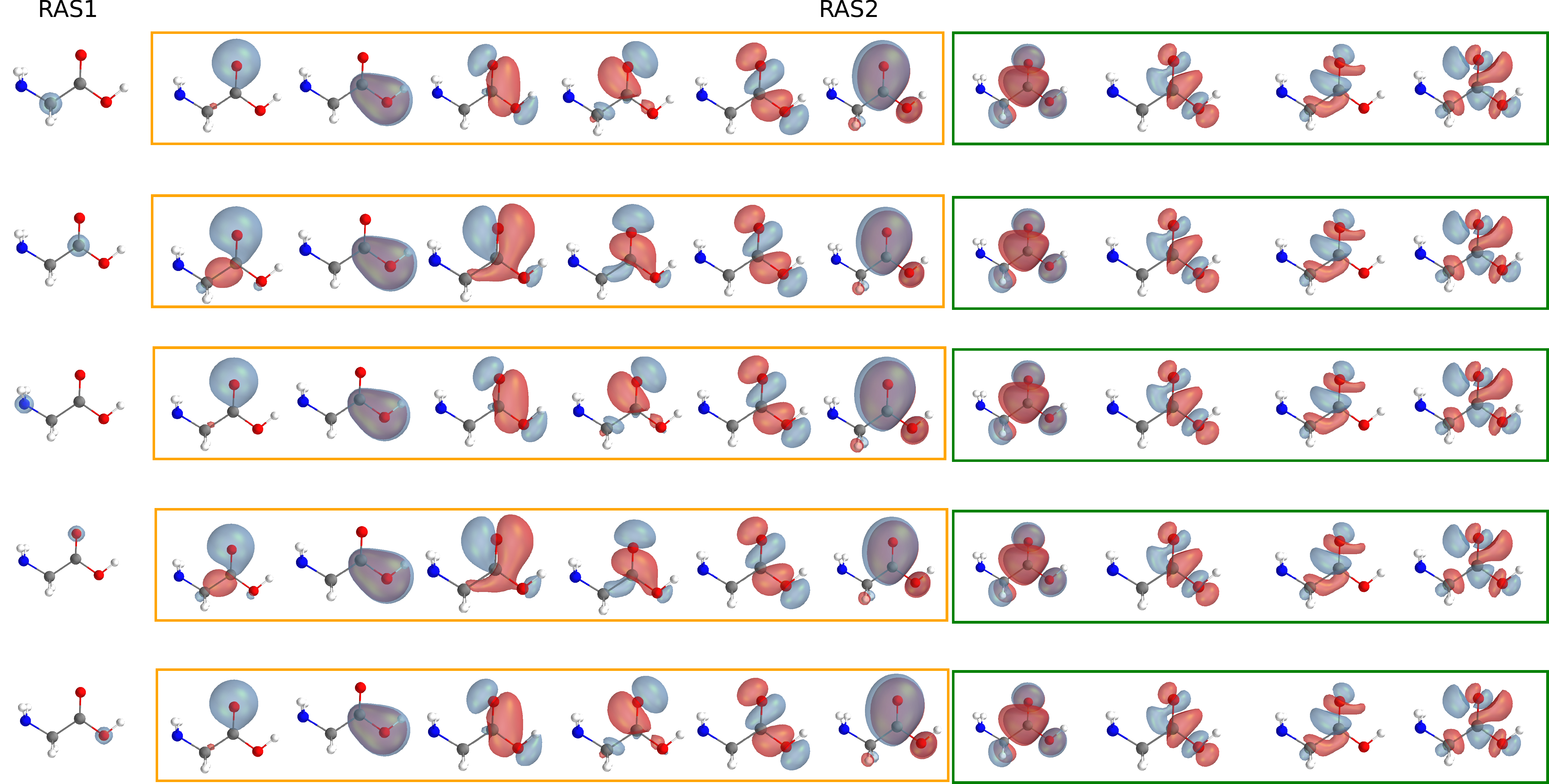}
    \caption{CASSCF natural orbitals corresponding to the active space (12,10) of glycine in the gas phase, shown separately for each nucleus: methylene carbon (C$_{H_2}$), carboxylic carbon (C$_O$), nitrogen (N), carbonyl oxygen (O$_C$), and hydroxyl oxygen (O$_H$). The orbitals are classified as above.}
    \label{fig:orb_vacuo_1210}
\end{figure}

\begin{figure}[ht!]
    \centering
    \includegraphics[width=0.9\linewidth]{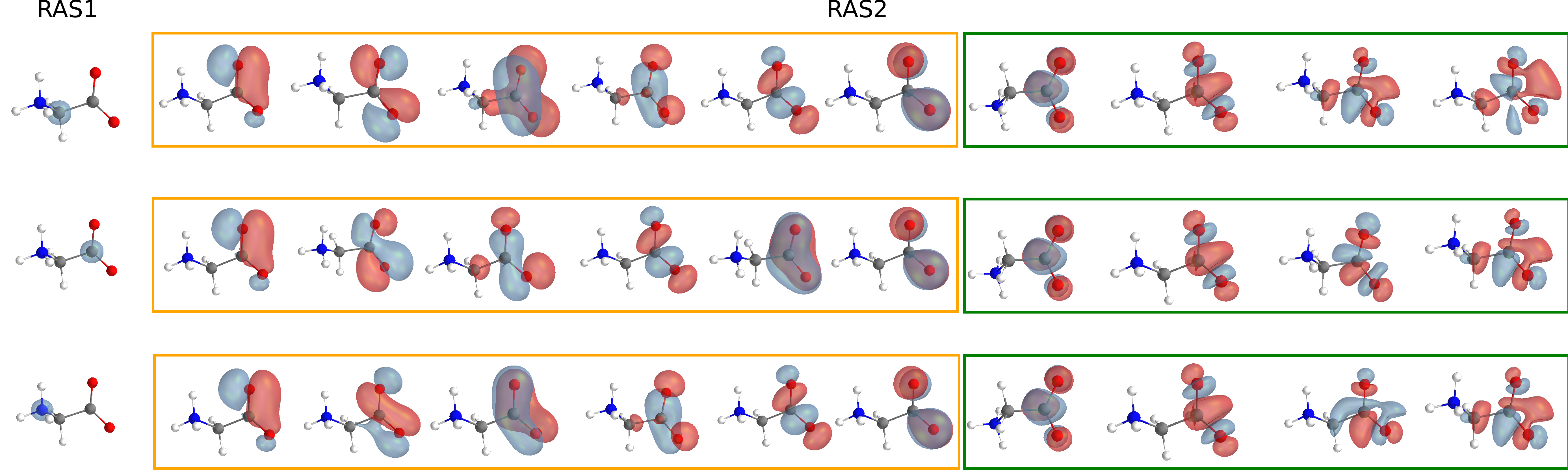}
    \caption{CASSCF natural orbitals corresponding to the active space (12,10) of a representative snapshot of glycine in aqueous solution, shown separately for each nucleus: methylene carbon (C$_{H_2}$), carboxylic carbon (C$_O$), nitrogen (N). The orbitals are classified as above.}
    \label{fig:orb_water_1210}
\end{figure}

\clearpage

\subsection{XPS Convergence}\label{sec:XPS_conv}

Although 200 snapshots were initially selected, the number of successfully converged electronic-structure calculations depends on the atomic edge, active space, and embedding parametrization. The ensemble sizes reported in the figure legends correspond to the configurations included in the final spectral averages.

\begin{figure}[ht!]
    \centering
    \includegraphics[width=0.8\linewidth]{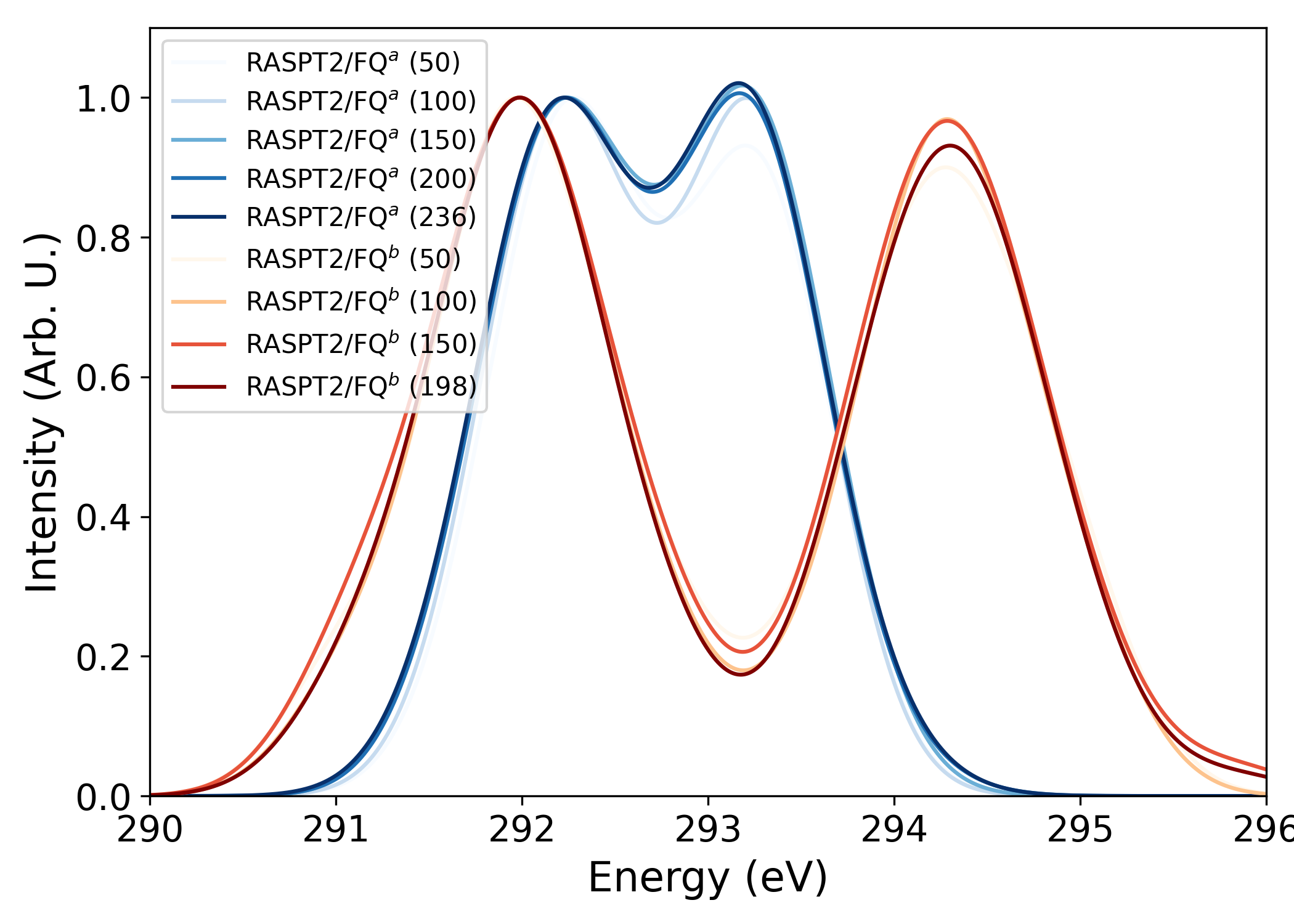}
    \caption{Convergence of RASPT2/FQ$^{(a,b)}$ C K-edge for the (12,7) active space with increasing number of spherical snapshots extracted from MD simulations.}
    \label{fig:c_conv_127}
\end{figure}

\begin{figure}[ht!]
    \centering
    \includegraphics[width=0.8\linewidth]{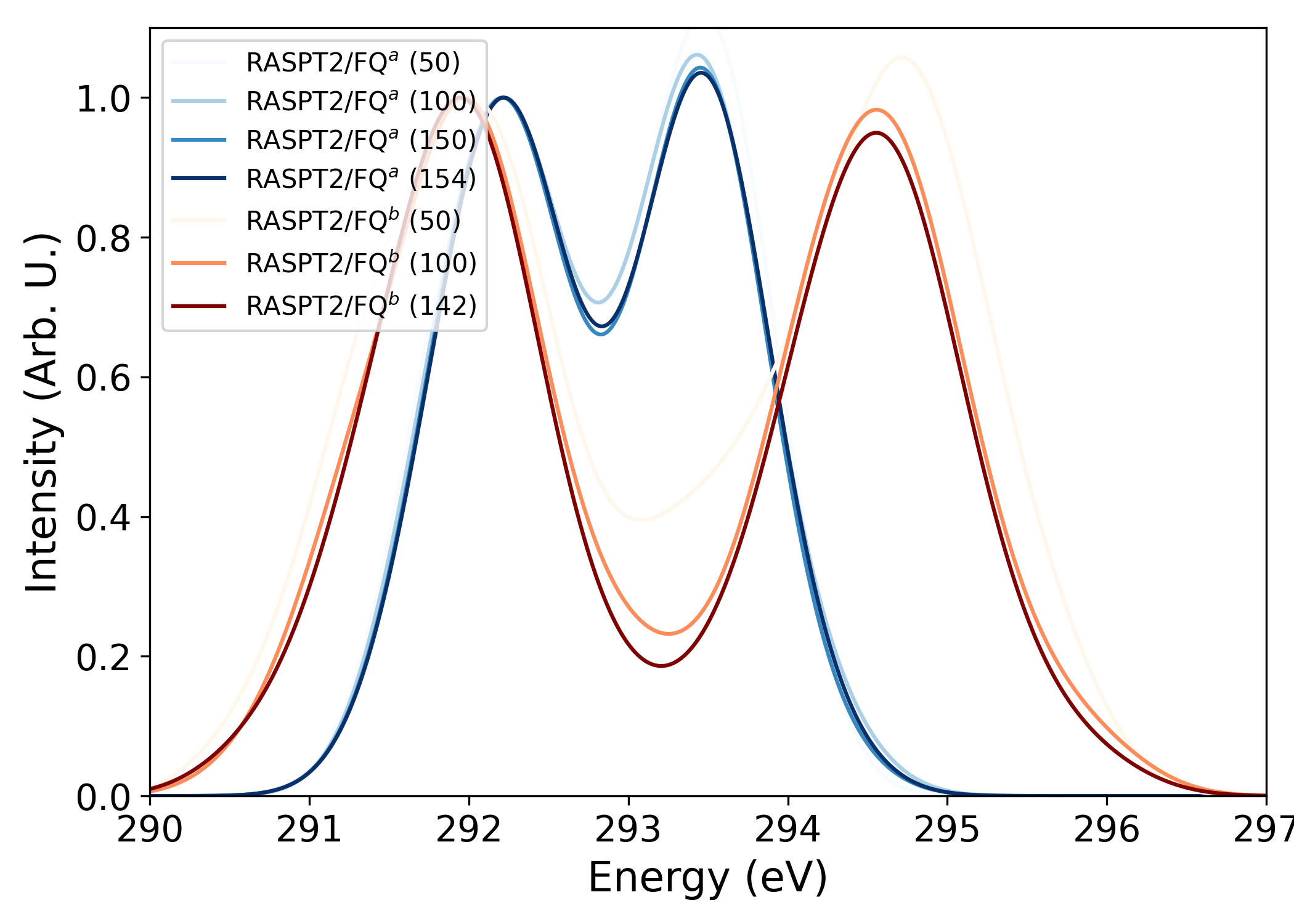}
    \caption{Convergence of RASPT2/FQ$^{(a,b)}$ C K-edge for the (12,10) active space with increasing number of spherical snapshots extracted from MD simulations.}
    \label{fig:c_conv_1210}
\end{figure}

\begin{figure}[ht!]
    \centering
    \includegraphics[width=0.8\linewidth]{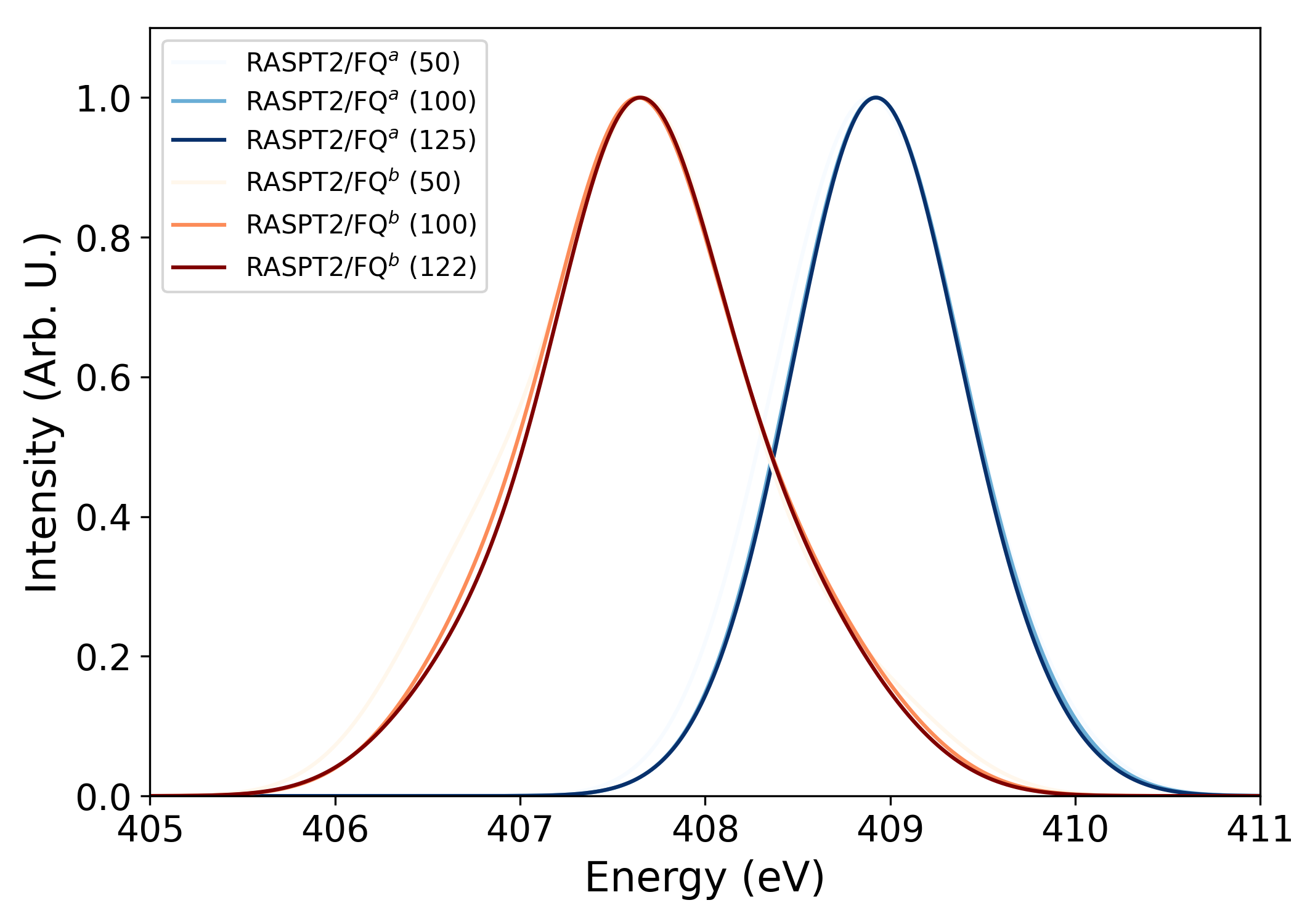}
    \caption{Convergence of RASPT2/FQ$^{(a,b)}$ N K-edge for the (12,7) active space with increasing number of spherical snapshots extracted from MD simulations.}
    \label{fig:n_conv_127}
\end{figure}

\begin{figure}[ht!]
    \centering
    \includegraphics[width=0.8\linewidth]{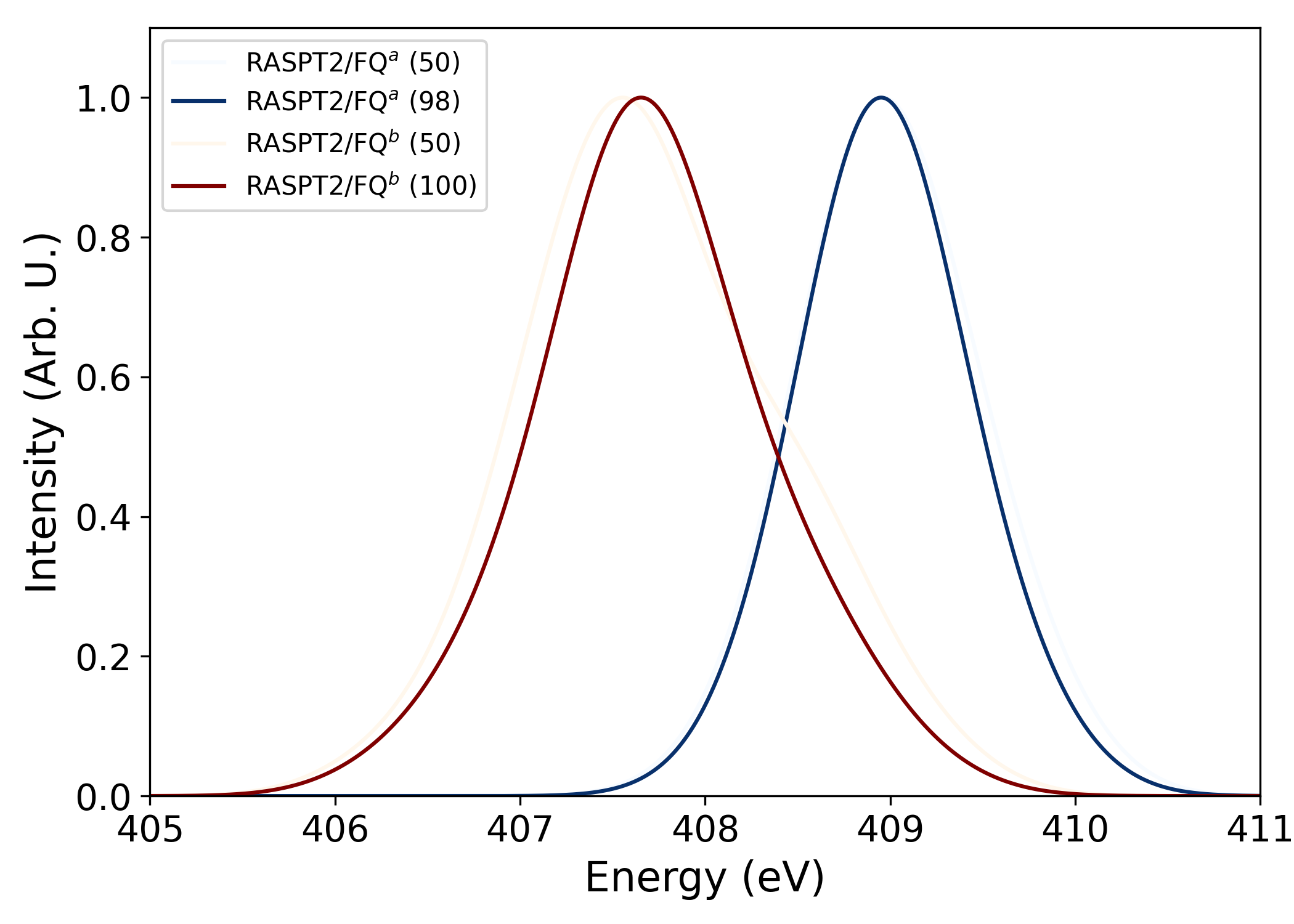}
    \caption{Convergence of RASPT2/FQ$^{(a,b)}$ N K-edge for the (12,10) active space with increasing number of spherical snapshots extracted from MD simulations.}
    \label{fig:n_conv_1210}
\end{figure}

\clearpage
\subsection{Additional gas-phase O K-edge XPS results}

\begin{figure}[ht!]
    \centering
    \includegraphics[width=0.8\linewidth]{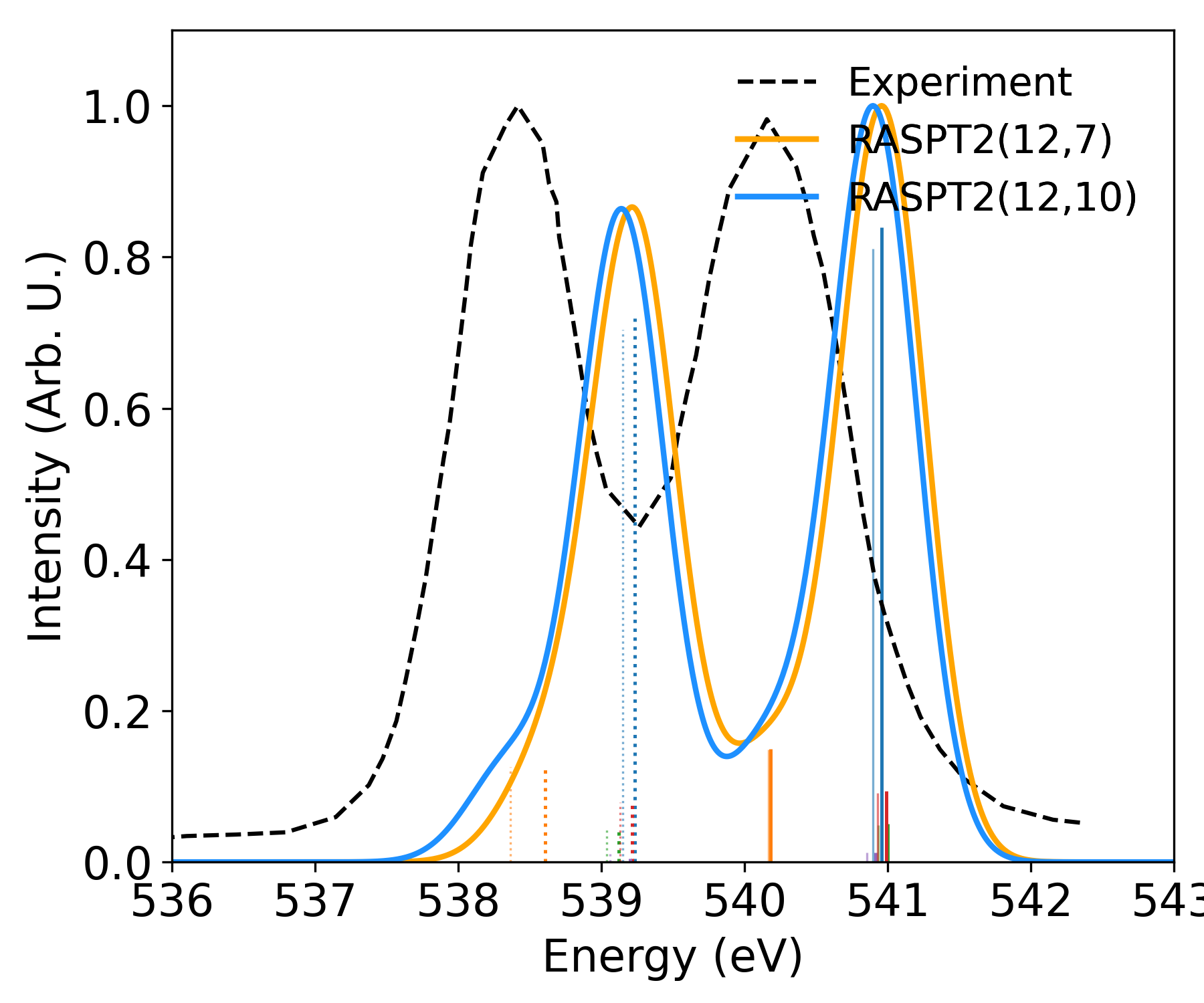}
   \caption{Gas-phase O K-edge XPS spectrum of glycine calculated at the RASSCF--RASPT2 level using the $(12,7)$ and $(12,10)$ active spaces. The corresponding site-resolved ionization energies are shown as stick spectra, while the experimental spectrum is reported as a dashed line. All calculations were performed using the ANO-RCC-VTZP basis set. Any rigid energy shifts applied to the calculated spectra are specified in the text.}
    \label{fig:o_xps_gas_phase}
\end{figure}

\clearpage

\section{XAS of glycine at C K-edge in gas-phase and aqueous solution}\label{sec:SI_xas_c}

\begin{figure}
    \centering
    \includegraphics[width=0.8\linewidth]{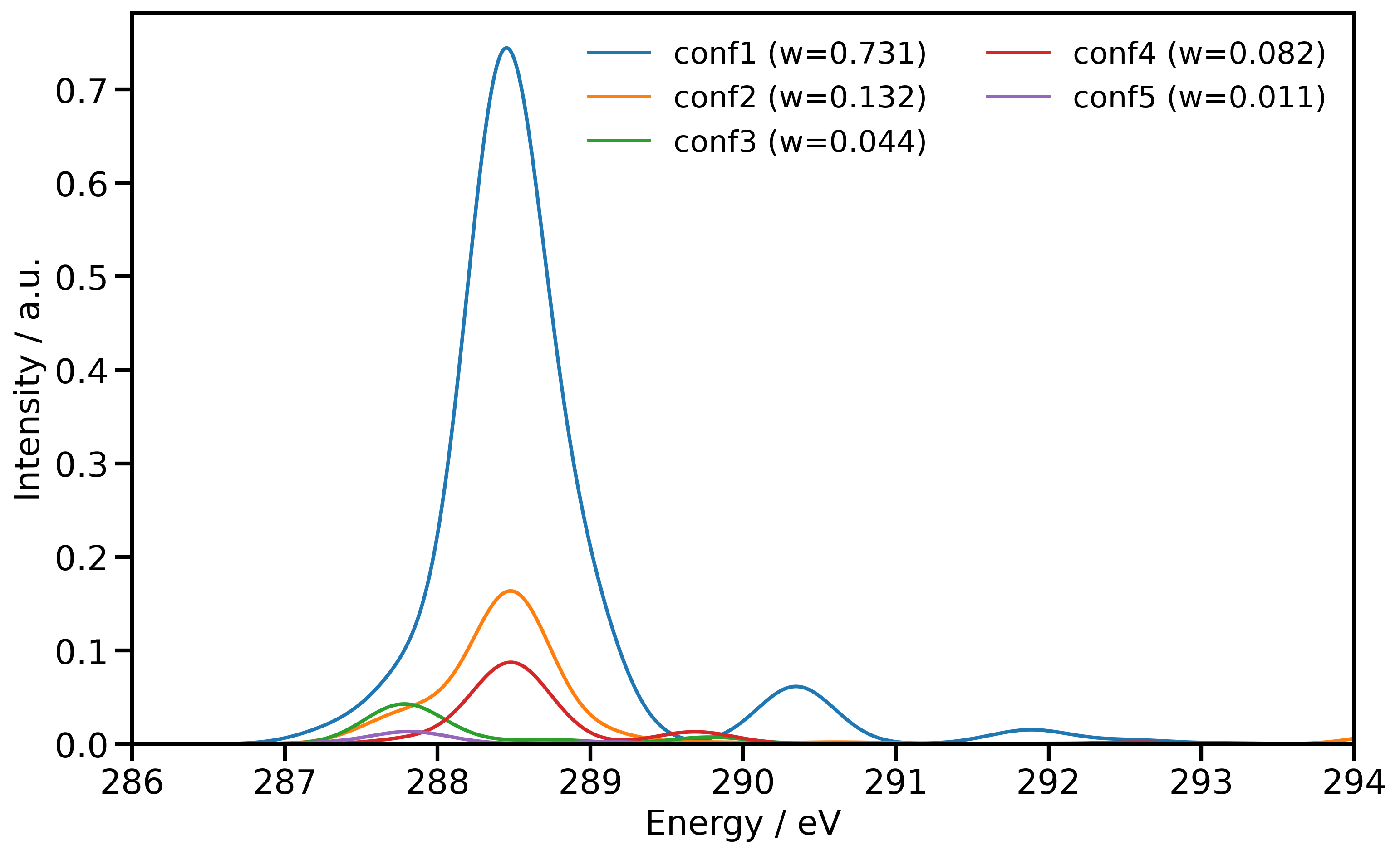}
    \caption{Conformational makeup of the gas-phase C K-edge spectrum of glycine}
    \label{fig:conform_C}
\end{figure}

\begin{figure}[H]
    \centering
    \includegraphics[width=0.99\linewidth]{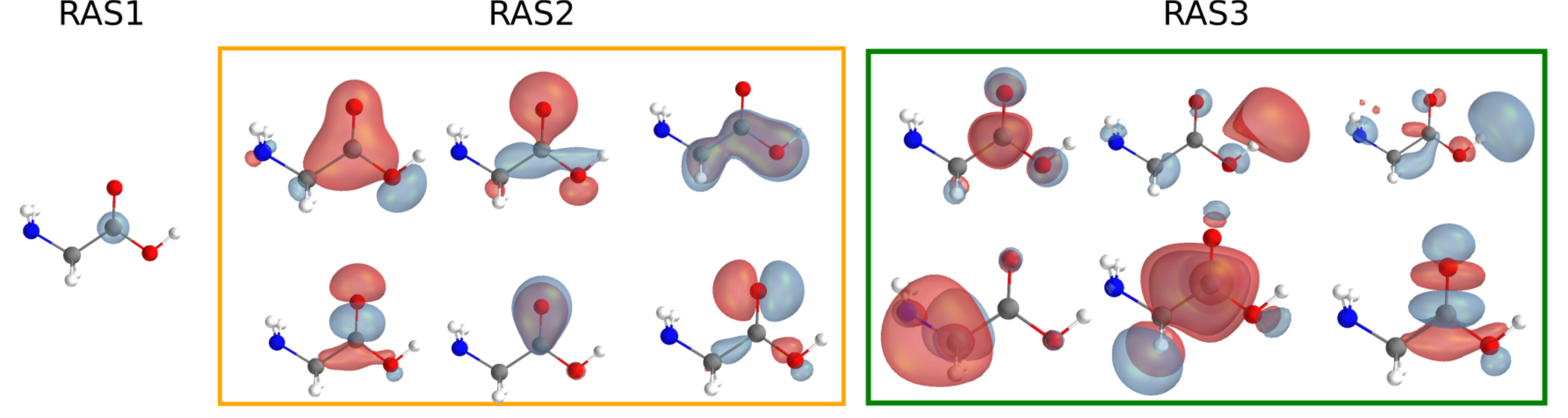}
    \caption{RASSCF natural orbitals for conformer I (see Ref. \cite{myhre2019x}) of gas-phase glycine obtained with the RAS1(1,1)/RAS2(12,6)/RAS3(2,6) (10 states) active space, corresponding to excitations from the C$_O$ core orbital.} 
    \label{fig:orbitals_co_gas}
\end{figure}

\begin{figure}[H]
    \centering
    \includegraphics[width=0.99\linewidth]{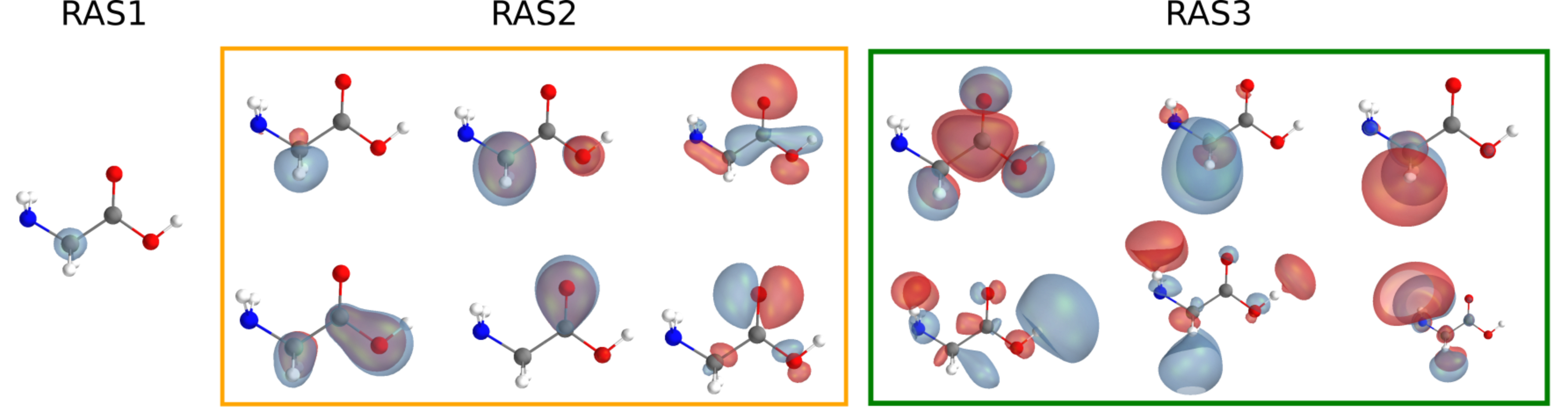}
    \caption{RASSCF natural orbitals for conformer I (see Ref. \cite{myhre2019x}) of gas-phase glycine obtained with the RAS1(1,1)/RAS2(12,6)/RAS3(2,6) (10 states) active space, corresponding to excitations from the C$_{H_2}$ core orbital.}
    \label{fig:orbitals_ch2_gas}
\end{figure}

\begin{figure}[H]
    \centering
    \includegraphics[width=0.99\linewidth]{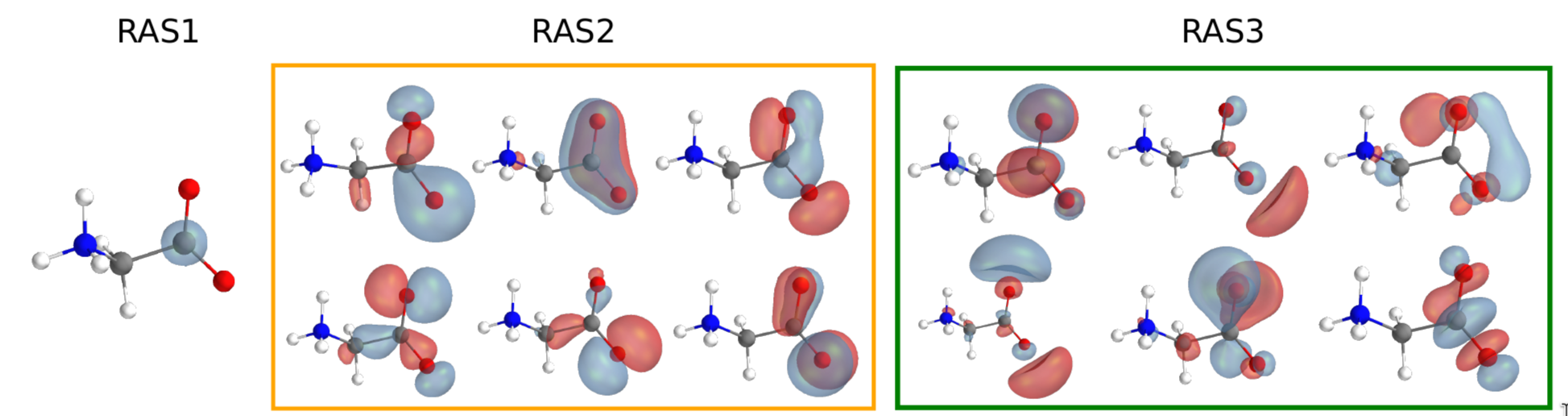}
    \caption{RASSCF natural orbitals for a representative snapshot in solution obtained with the RAS1(1,1)/RAS2(12,6)/RAS3(2,6) (10 states) active space, corresponding to excitations from the C$_O$ core orbital.}
    \label{fig:orbitals_co_water}
\end{figure}

\begin{figure}[H]
    \centering
    \includegraphics[width=0.99\linewidth]{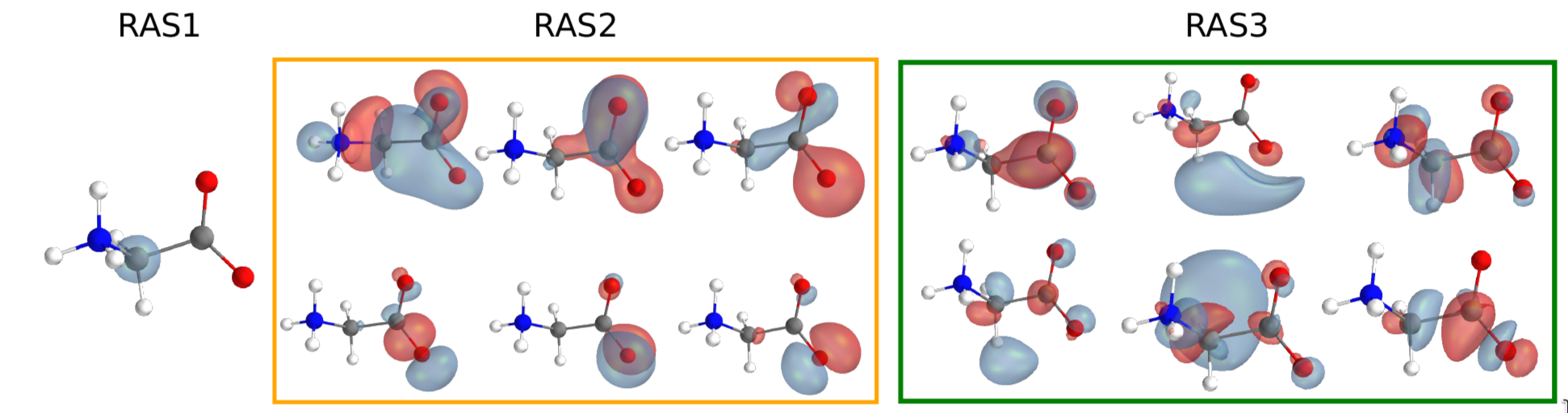}
    \caption{RASSCF natural orbitals for a representative snapshot in solution obtained with the RAS1(1,1)/RAS2(12,6)/RAS3(2,6) (10 states) active space, corresponding to excitations from the C$_{H_2}$ core orbital.}
    \label{fig:orbitals_ch2_water}
\end{figure}

\section{XAS of glycine at N K-edge in gas-phase and in aqueous solution}\label{sec:SI_xas_n}

\begin{figure}[H]
    \centering
    \includegraphics[width=0.8\linewidth]{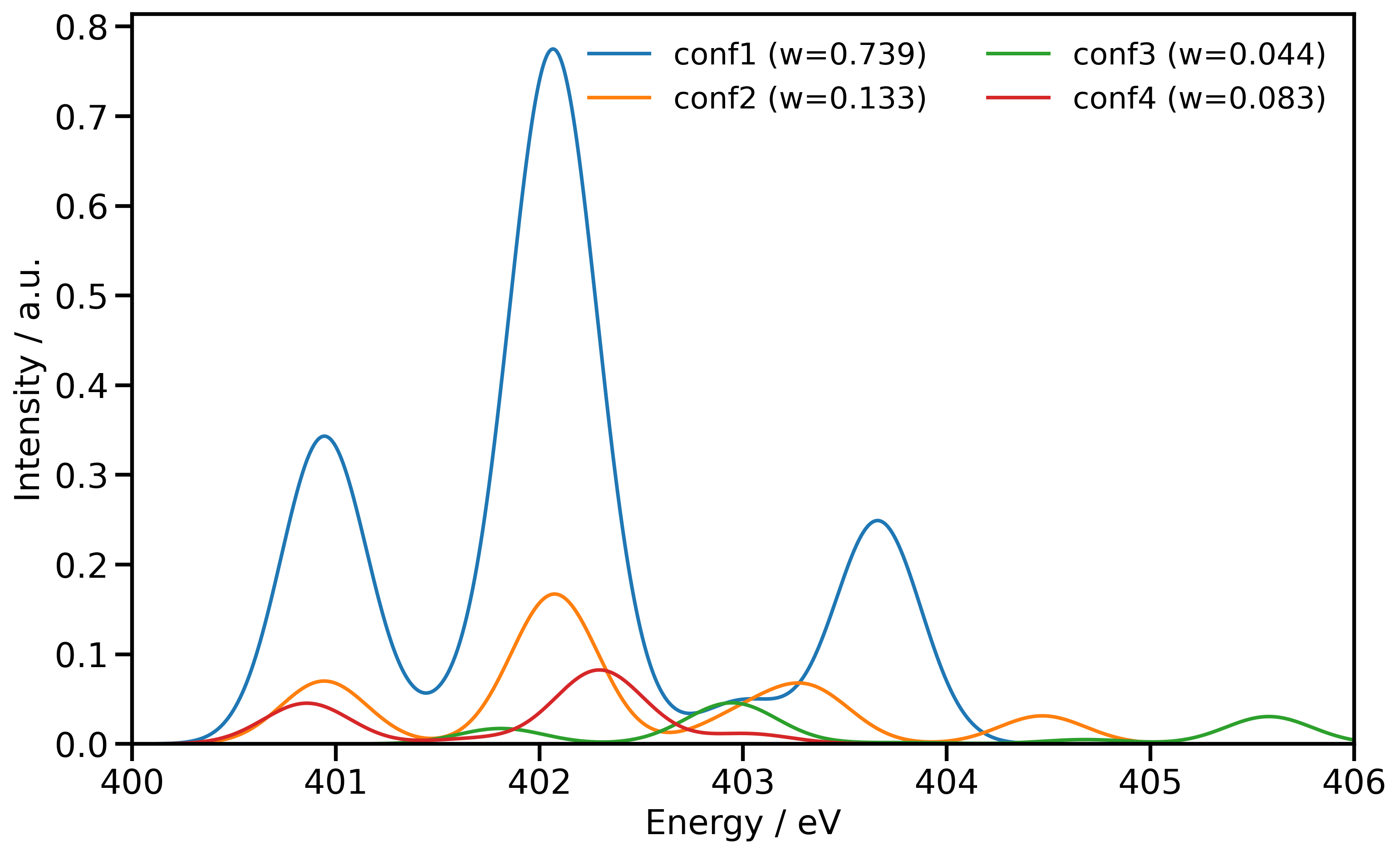}
    \caption{Conformational makeup of the gas-phase N K-edge spectrum of glycine}
    \label{fig:conform_N}
\end{figure}

\begin{figure}[hb!]
    \centering
    \includegraphics[width=0.99\linewidth]{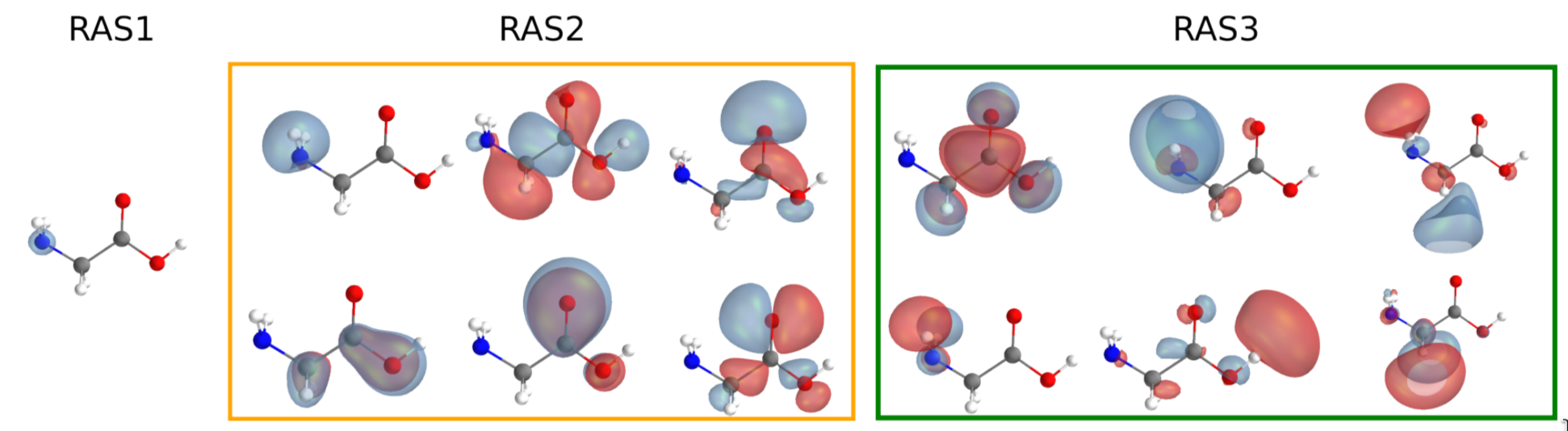}
    \caption{RASSCF natural orbitals for conformer I (see Ref. \cite{myhre2019x}) of gas-phase glycine obtained with the RAS1(1,1)/RAS2(12,6)/RAS3(2,6) (10 states) active space, corresponding to excitations from the N core orbital.}
    \label{fig:orbitals_n_gas}
\end{figure}

\begin{figure}[ht]
    \centering
    \includegraphics[width=0.99\linewidth]{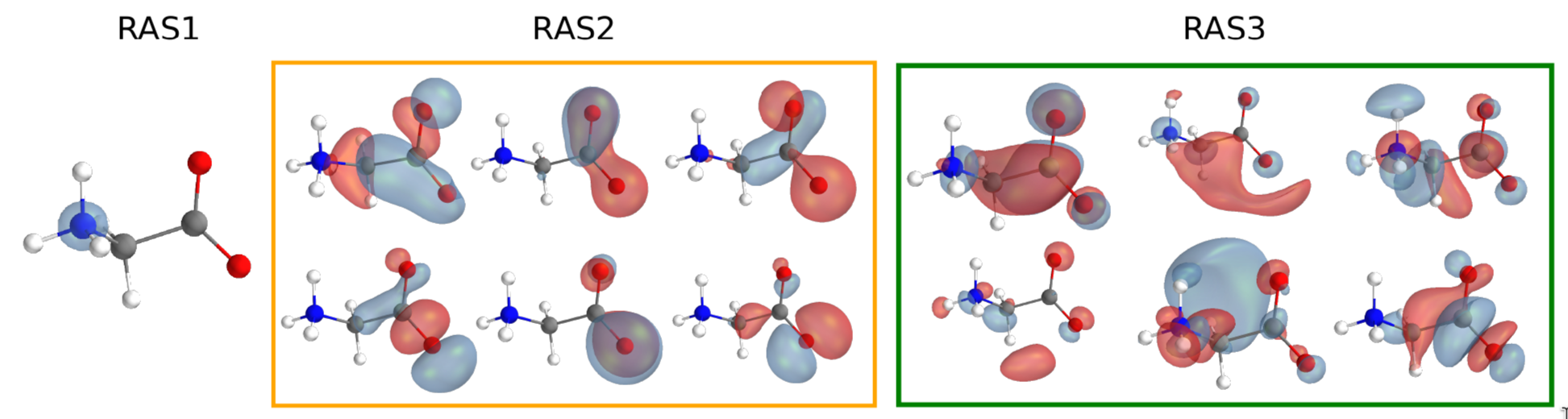}
    \caption{RASSCF natural orbitals for a representative snapshot in solution obtained with the RAS1(1,1)/RAS2(12,6)/RAS3(2,6) (10 states) active space, corresponding to excitations from the N core orbital.}
    \label{fig:orbitals_n_water }
\end{figure}

\section{XAS convergence}\label{sec:SI conv}

\begin{figure}[H]
    \centering
    \includegraphics[width=0.99\linewidth]{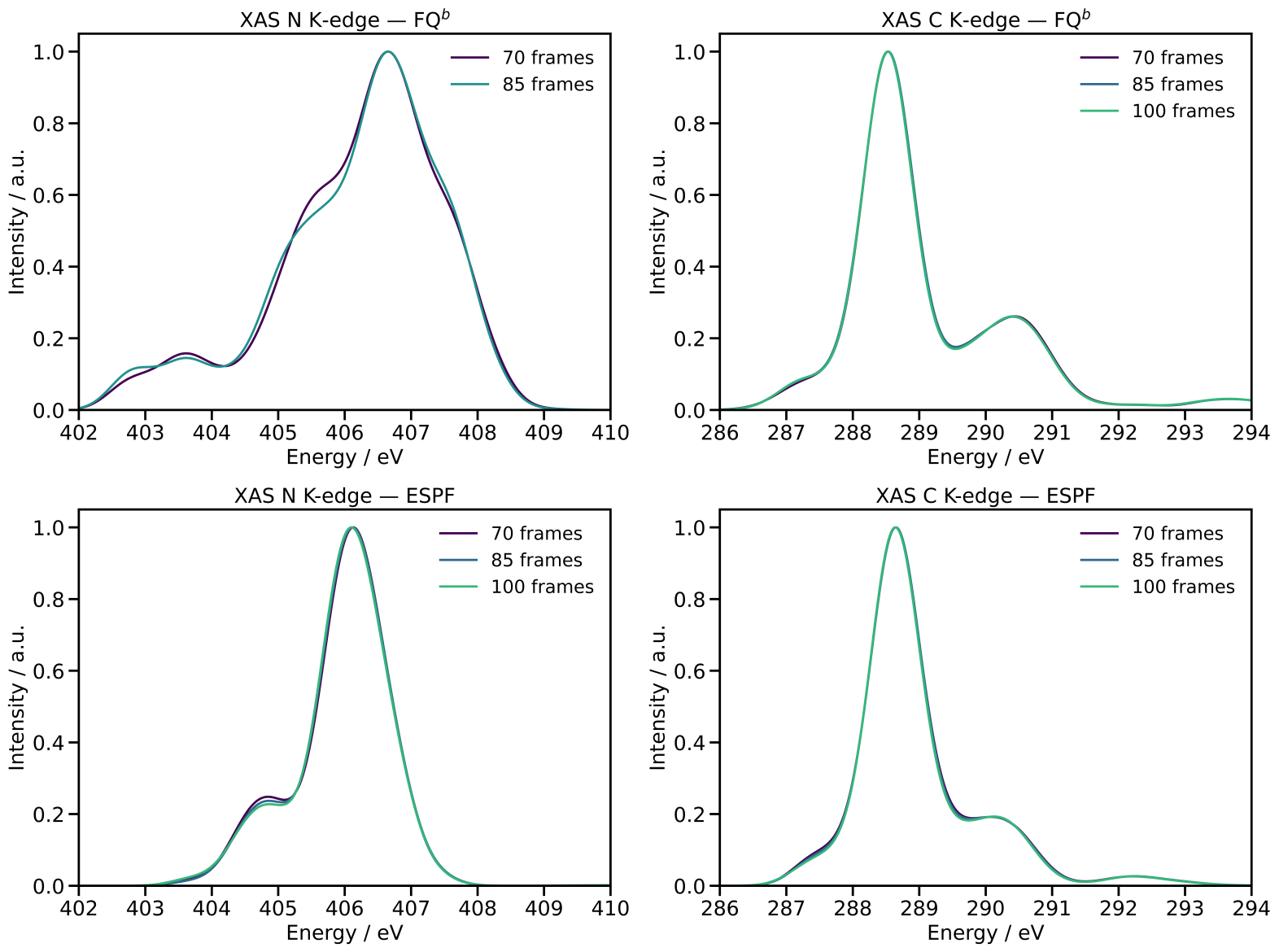}
    \caption{Convergence of the XAS spectra in the two different solvation models, FQ$^b$ (top) and ESPF (bottom) as well as for both edges, N (left) and C (right).}
    \label{fig:xas_conv}
\end{figure}

%% file: main.bbl
\begin{mcitethebibliography}{134}
\bibitem[Lin and Truhlar(2007)Lin, and Truhlar]{lin2007qm}
Lin,~H.; Truhlar,~D.~G. QM/MM: what have we learned, where are we, and where do we go from here? \emph{Theor. Chem. Acc.} \textbf{2007}, \emph{117}, 185--199\relax
\mciteBstWouldAddEndPuncttrue
\mciteSetBstMidEndSepPunct{\mcitedefaultmidpunct}
{\mcitedefaultendpunct}{\mcitedefaultseppunct}\relax
\EndOfBibitem
\end{mcitethebibliography}
